\documentclass[letter]{aastex631}
\usepackage[utf8]{inputenc}

\usepackage{graphicx}	
\usepackage{amsmath}	
\usepackage{amssymb}	
\usepackage{hyperref}

\begin{document}

\title{Hybrid Emission Modeling of GRB 221009A: Shedding Light on TeV Emission Origins in Long-GRBs}


\author[0000-0002-8422-6351]{Hebzibha Isravel}
\affiliation{Ben-Gurion University of the Negev \\
Beer-Sheva 8410501, Israel}

\author[0000-0003-4477-1846]{Damien B\'egu\'e}
\affiliation{Bar-Ilan University\\
Ramat-Gan 5290002, Israel}

\author[0000-0001-8667-0889]{Asaf Pe'er}
\affiliation{Bar-Ilan University\\
Ramat-Gan 5290002, Israel}

\begin{abstract}
Observations of long duration gamma-ray bursts (GRBs) with TeV emission during their afterglow have been on the rise. Recently, GRB 221009A, the most energetic GRB ever observed, was detected by the {LHAASO} experiment in the energy band 0.2 - 7 TeV.
Here, we interpret its afterglow in the context of a hybrid model in which the TeV spectral component is explained by the proton-synchrotron process while the low energy emission from optical to X-ray is due to synchrotron radiation from electrons. We constrained the model parameters using the observed optical, X-ray and TeV data. By comparing the parameters of this burst and of GRB 190114C,  
we deduce that the VHE 
emission at energies $\geq$ 1 TeV in the GRB afterglow requires large explosion 
kinetic energy, $E \gtrsim 10^{54}$~erg and a reasonable
circumburst density, $n\gtrsim 10$~cm$^{-3}$. This results in a small injection fractions 
of particles accelerated to a power-law, $\sim 10^{-2}$. {A  significant fraction of shock energy must be allocated to a near equipartition magnetic field, $\epsilon_B \sim 10^{-1}$, while electrons should only carry a small fraction of this energy, $\epsilon_e \sim 10^{-3}$. Under these conditions required for a proton synchrotron model, namely  $\epsilon_B \gg \epsilon_e$, the SSC component is substantially sub-dominant over proton-synchrotron as a source of TeV photons.}
These results lead us to suggest that proton-synchrotron process is a strong contender for the radiative mechanisms explaining GRB afterglows in the TeV band. 
\end{abstract}


\section{Introduction}\label{sec:Intro}
Gamma Ray Bursts (GRB) are indisputably the universe's 
brightest extragalactic transient events. They feature a brief 
prompt phase emission, mostly observed in the energy extending from a few keVs to a few GeVs. It is then followed by the extended broadband afterglow, detected at all energy bands, from radio
to a few hundreds of GeVs and possibly even higher \citep[for reviews see \textit{e.g.}][]{1999PhR...314..575P, 2006RPPh...69.2259M, 2015PhR...561....1K, zhang_2018}. 
Detections of GRB afterglows at the highest 
energies (i.e., $>300$ GeV) have been on the rise in the past two 
decades \citep[for reviews, see, e.g.,][]{2018IJMPD..2742003N, 2022Galax..10...66M}.
Understanding the physics underlying this emission has become
of highest importance as it holds the clues to better constrain and understand the afterglow of GRBs.

The recent development of highly sensitive
ground-based detectors, such as the High Energy Stereoscopic System \citep[H.E.S.S, ][]{1997APh.....6..369A, 2021Sci...372.1081H},
the Major Atmospheric Gamma Imaging Cherenkov \citep[MAGIC, ][]{lorenz_magic_2005}, and 
the more recent Large High Altitude Air Shower Observatory 
\citep[{LHAASO}, ][]{2019arXiv190502773C}, has allowed for the detection of sub-TeV to 
$\sim$ TeV signal in GRBs and measurements of their spectra in this band.
Examples include GRB 180720B \citep{2019Natur.575..464A}, GRB 190114C 
\citep{acciari_observation_2019}, GRB 190829A \citep{2021Sci...372.1081H} and GRB 201216C \citep{2020GCN.29075....1B}. 

The emission mechanism of GRBs with emission at the very high energy (VHE, $\geq$ TeV) band is highly debated.  With reference to the fireball scenario, 
the synchrotron self-Compton (SSC) model is 
prominent among the radiation mechanisms that aim to explain these
signals. In this mechanism, photons emitted by the synchrotron process at low energies, are upscattered to VHE by the energetic electrons that emitted them 
{\citep{ghisellini1998quasi, Dermer_2000, 2001ApJ...548..787S,fraija2019synchrotron, 2019ApJ...884..117W,  Derishev:2021ivd, 2022ApJ...934..188F, 2022MNRAS.512.2142Y}}. Alternatively, the proton-synchrotron mechanism has 
been suggested to produce a photon signal at these extreme energies
 \citep{PhysRevLett.78.4328, 1998ApJ...499L.131B, 2022arXiv221002363I, 2022arXiv221105754Z}.
 The idea is that the same mechanism responsible for accelerating electrons also accelerates protons to high energies. The energetic protons then emit the observed VHE photons,  
 reaching energies as 
high as $\gtrsim$~TeV \citep{1998ApJ...totania, 2001ApJ...559..110Z, 
2022arXiv221002363I}.
In addition, photo-pion and photo-pair 
production processes may also produce VHE photons  
\citep[see e.g.][]{2010OAJ.....3..150R} although this requires a compact region, and it is not clear how this is obtained at late times.

The recent detection of TeV photons from GRB 221009A allows for the first time to probe in great details the GRB afterglow phase in this energy band.
This GRB is by far the brightest GRB ever detected \citep{2023arXiv230314172L}. It was observed at the energy band 0.2 - 7 TeV by LHAASO  \citep{doi:10.1126/science.adg9328}. 
The isotropic equivalent luminosity in the band $0.3-5$~TeV is $7.3 \times 10^{50}$~erg~s$^{-1}$ and the observed peak flux is $\sim 1.2 \times 10^{-5}$~erg~cm$^{-2}$~s$^{-1}$ \citep{doi:10.1126/science.adg9328}.
This GRB is a nearby burst, at a cosmological redshift $z = 0.151$ 
\citep{web:nasa:redshift, 2022GCN.32686....1C} as well as the brightest burst ever detected, with an isotropic equivalent burst 
energy $E_{iso} \simeq 3 \times 10^{54}$~erg \citep{web:nasa:isotro}. The half opening angle of the jet is estimated to be $\sim 0.8^\circ$ \citep{doi:10.1126/science.adg9328}.

 Several authors considered the conventional SSC model to interpret the VHE afterglow spectrum of GRB 221009A. This process is a natural outcome of the classical synchrotron-SSC emission model, and can account for emission in this energy band. However, it is not clear yet whether this model can explain the broad-band data (at all wavelengths), given the strong constraints on the TeV band flux from  radio, optical and X-ray data \citep{2022arXiv221015857G, 2022Galax..10...66M}. Furthermore, this model cannot explain a $\gtrsim$ 10 TeV energy photons \citep{web:nasa:LHAASO} originally claimed to be observed \citep{ 2022arXiv221015857G, 2022arXiv221010673R, 2023arXiv230206225K, 2023A&A...670L..12D, 2023arXiv230204388L}. On the other hand, it is not clear if photons at these energies were detected \footnote{In their analysis, the {LHAASO} collaboration cut the spectrum at 7~TeV.} and required to explain the observed spectra \citep{doi:10.1126/science.adg9328}.
A recent work by \citet{2022arXiv221105754Z}  considered the possibility that proton-synchrotron 
may be the source of  $\gtrsim$ TeV energy photons in the reverse shock scenario, and concluded that this is a plausible scenario under certain conditions, in particular a very strong magnetic field. 

Here, we use the data available 
from optical through X-rays to TeV, and show that
the synchrotron emission from relativistic protons
can explain both the flux and the temporal features
of the VHE afterglow of GRB 221009A, while its lower-energy afterglow
counterpart is interpreted with the electron-synchrotron process.
We determined two sets of parameters able to explain the observational features of this burst.
Then by comparing these model parameters with those deduced
for GRB 190114C \citep{2022arXiv221002363I}, we identify a set of consistent 
characteristics for the VHE afterglows with energies $\gtrsim$ TeV, within the framework of the hybrid model we present.

This paper is structured as follows. In section 
\ref{sec:Data_Analys} we review the available 
data on GRB 221009A obtained by various
space-based and ground-based facilities. 
In section \ref{sec:Model}, we present our model within the context
of the standard fireball scenario.
We then use the data to constrain the values of the free
physical parameters in section \ref{sec:Limits}.  The SEDs are then produced for three different cases in section \ref{sec:Results}. 
We investigate the common features 
encountered in the VHE afterglows of GRBs in section \ref{sec:discussion}. 
Finally, our conclusions follow in section \ref{sec:conclusion}.

\section{Observational Data of GRB 221009A}
\label{sec:Data_Analys}

The long-duration GRB 221009A triggered the
Gamma-Ray Burst Monitor (GBM) on board the  \textit{Fermi}  spacecraft
on October 9, 2022, at $T_0 =$~UT 13:16:59 \citep{web:nasa:FermiGBM}.
Initially, the GBM captured two separate emission episodes
\citep{2022GCN.32642....1L}. The first occurred between $T_0$-0 and $T_0$+43.4 s
with a reported peak energy of $375\pm 87$~keV and a fluence of $2.12 \pm 0.05 \times 10^{-5}$~ erg cm$^{-2}$ in the energy range 10-1000~keV.
The second episode, being the brightest, exhibited numerous peaks during
the time interval $T_0$+175 to $T_0$+1458~s. Due to the saturation of the
detectors caused by the accumulation of photons in several of these peaks, the 
exact flux can hardly be measured. Yet, the KONUS-WIND collaboration recently 
reported the fluence $\sim 0.21~ \rm erg~ \rm cm^{-2}$ within 
the energy band 20 keV - 10 MeV \citep{2023arXiv230213383F}. 

{The High Energy (HE) X-ray telescope on board the \textit{Insight}-Hard X-ray Modulation Telescope (\textit{Insight}-HXMT) also 
triggered and monitored this burst on 9$^{th}$ October 2022, at 13:17:00.050 UT \citep{2022ATel15660....1T}. 
This instrument's primary goal is to observe GRBs and electromagnetic counterparts of gravitational waves \citep{Cai:2021lmr}. The \textit{Insight}-HXMT together with the Gravitational Wave High-energy Electromagnetic Counterpart All-sky Monitor (GECAM-C) measured the emission in the energy band $\sim$10~KeV to 6 MeV starting from the precursor of the event until the early afterglow phase for a duration of about $\sim$1800~s  \citep{2023arXiv230301203A}. It was determined that the burst has a total isotropic energy of $\approx 1.5 \times 10^{55}$ ergs. }

The \textit{Fermi-}Large Area Telescope (LAT)  subsequently
observed this GRB 
between 200 and 800~s following the GBM trigger
\citep{web:nasa:FermilAT}. It is the brightest GRB
ever detected by LAT, with a maximum reported photon energy of 99.3~GeV,
observed 240~s after $T_0$. Due to the extreme brightness, the \textit{Fermi-}-LAT 
detector was saturated during the time period 200-400 s \citep[corresponding to
"bad" time intervals, where the exact flux could not be measured due
to the saturation; see][]{web:nasa:LAT1, web:nasa:LAT2}. 
 The LAT data
in the energy band 0.1 -1 GeV between 400~s and 800~s
was modeled by a power-law spectrum
${dN}/{dE} = N_0 \left({E_{LAT}}/{E_f}\right)^{p_0}$ resulting
in a spectral index $p_0 = 1.87\pm 0.04$  and in a photon flux of
$\Phi_{\gamma} = 6.2 \pm 0.4 \times  10^{-3}$~ ph cm$^{-2}$ s$^{-1}$\citep{web:nasa:FermilAT}.

Nearly 53.3 minutes after the GBM trigger, at UT 14:10:17, the 
\textit{Swift}-Burst Alert Telescope (BAT) also triggered and observed GRB 221009A in the hard X-ray band
\citep{2022ATel15650....1D}. Starting 143 s after the BAT trigger,
\textit{Swift}-XRT slewed and monitored the then steadily declining X-ray light curve with a photon index $1.836\pm 0.012$ and a temporal index of $1.509 \pm 0.004$ \citep{2007A&A...469..379E,2009MNRAS.397.1177E}.

The optical afterglow in the R-band was measured at 18:45 UT,
4.6 hours after the BAT trigger, corresponding to 5.5 hours after
the GBM trigger, with
magnitude 16.57 $\pm$ 0.02 by the Observatiorio Sierra Nevada (OSN) in Spain \citep{web:nasa:Rband}. 
Considering the strong galactic extinction in the R-band, the AB magnitude is estimated to be 3.710 \citep{1998ApJ...500..525S}. The optical and infrared data between 0.2 and 0.5 days are presented in \citet{2023arXiv230207906O} and \citet{2023arXiv230414331G}, and are corrected for the galactic extinction of 1.32 mag. For instance, 
179 s after the BAT trigger, \textit{swift-}UVOT observed 
GRB 221009A and recorded a magnitude of 16.68 $\pm$ 0.03 in the white filter \citep{2022GCN.32656....1K}.

Finally, LHAASO's Water Cheronkov Detector Array (WCDA) \citep{2019arXiv190502773C} observed GRB 221009A within its field of view at the time of the GBM trigger. 
Within a span of 3000~s from the burst trigger, more than 60,000 photons in the energy band 0.2 $\sim$ 7  TeV were detected by the LHAASO \citep{doi:10.1126/science.adg9328}. 
Around the phase of the main burst, LHAASO recorded the flux of $\sim 6 \times 10^{-8}$~erg cm$^{-2}$ s$^{-1}$ at 1 TeV at the time period between $T_0 + 220 s - T_0 + 230 s$, after correcting for extra-galactic background light (EBL)  attenuation\footnote{As the VHE photons traversing through cosmological sources experience $\gamma\gamma 
-$ pair-production by interacting with the EBL, which substantially attenuates the intrinsic spectrum of the source \citep{doi:10.1126/science.1227160}.}.

The {LHAASO} collaboration and the Fermi-GBM collaboration deduced different times for the onset of the afterglow. 
\citet{2023arXiv230314172L} for the Fermi-GBM collaboration argued that the beginning of 
the afterglow phase was $\sim 597$~s after the trigger. This is based on the inability of a single decay function to explain the lightcurve at earlier times. On the other hand, 
interpretation of the {LHAASO} data in the framework of the external shock, 
based on the temporal decay of the light-curve lead to estimating the onset of the afterglow at this band already at 226~s after the GBM trigger \citep{doi:10.1126/science.adg9328}.
The origin of this discrepancy can be due to the superposition of both prompt signal (which should originate from a small radius) 
and afterglow signal (originating from a forward shock propagating ahead of the jet, at larger radius) in the observations around few hundreds seconds. 
Therefore, here we will model the {LHAASO} emission as part of the afterglow and will not attempt to model the GBM data, which, as we will show below, is much brighter than the predicted GBM flux within the framework of our model (assumed to be produced by electron-synchrotron in this energy band).

\section{Model Description}
\label{sec:Model}

In this section, we detail
the afterglow dynamics and the emission mechanisms, which serve as
the basis of our model attempting to explain the VHE observation of GRB
221009A. We set our analysis within the framework 
of the fireball evolution scenario \citep{1990ApJ...348..485P,Piran:1993b,
1998ApJ...499..301M}, further assuming that the high-energy component (GeV
$\lesssim E_\gamma \leq$ TeV) and the low-energy component (eV
$\lesssim E_\gamma \leq$ MeV) of the observed spectrum are produced
by synchrotron radiation from the accelerated protons and
electrons, respectively via the external shock acceleration. More details on the processes and the model can be
found in \citet{2022arXiv221002363I}, and we remind here
only the key assumptions and equations. 

When the relativistic jet originating
from the compact GRB progenitor
encounters the stationary ambient environment,
 an outward propagating shock-wave is
created \citep{1993ApJ...418L...5P, 1999ApJ...526..697M}. This
shock collects and accelerates the ambient matter (both protons and electrons)
and generates in-situ a magnetic field. The accelerated 
particles then produce the observed multi-wavelength emission
\citep[see \textit{e.g.}][]{Sari:1995ApJ, Sari_1998, 2000ApJ...543...66P}. 
During the afterglow phase, the emission occurs while
the outflow expands in a self-similar way, following the
\citet{Blandford:1976} solution. We assume here that the
ultra-relativistic expansion can be considered adiabatic, \textit{i.e.}
that the radiative losses of the plasma behind the shock are negligible.
This is a good approximation for our scenario as accelerated protons
should carry most of the internal energy while they do not radiate efficiently.

Under those assumptions, the Lorentz factor of the jet, at
a given observed time $t$, is determined only by the isotropic-equivalent
explosion kinetic energy $E$, and the ambient ISM density $n$: 
\begin{align}\label{Eq:Gamma}
   \Gamma(E, n; t) = \left[ \frac{17 E (1+z)}{1024 \pi n m_p c^5 t^{3}} \right]^{1/8} = 61.3 \hspace{0.3em} {E_{54}^{1/8}}{ n_0^{-1/8}} t_3^{-3/8}, 
\end{align}
where $c$ is the speed of light, $m_p$ is the mass of the proton and we
took the redshift to be $z=0.151$ relevant for GRB 221009A. Here and below, $Q = 10^xQ_x$ in
cgs units is employed. Using $t = r/(4 \Gamma^2(r) c)$, one can express
the location of the blast wave as a function of the observed time, 
\begin{equation}
    r(E, n; t) = \left[\frac{17 E t }{4 \pi n m_p c (1+z)}\right]^{1/4} = 3.9 \times 10 ^{17}~ {E_{54}^{1/4} }{n_0^{-1/4} t_{3}^{1/4}  }\quad\text{cm}.
\end{equation}
Finally, we define the comoving shock expansion (dynamical) time 
as $t_{dyn} = r/ 
(\Gamma c) = 2.1 \times 10^5~ {E_{54}^{1/8} t_{3}^{5/8}}{n_0^{-1/8}}$~s.

In order to estimate the observed spectrum, we need to specify the magnetic field and the particle distribution functions. For the former, we take the standard assumption that an (uncertain) fraction, $\epsilon_B$,
of the post-shock thermal energy is used in generating a magnetic field. This gives
{
\begin{align}\label{Eq:B}
B = \sqrt{32\pi \epsilon_B \Gamma^2 n m_p c^2} = 7.5 \hspace{0.3em} {E_{54}^{1/8} \epsilon_{B,-1}^{1/2}n_0^{3/8}}{t_{3}^{-3/8}}\text{~G}. 
\end{align}
}
For the radiating particles, namely protons and electrons, we
assume that a fraction $\xi_x$ of all the particle is injected in the
radiative zone with a power-law distribution between some minimum Lorentz
factor $\gamma_{m, i}$ and maximum Lorentz factor, $\gamma_{\rm max, i}$, such that they
carry a fraction $\epsilon_i$ of the available internal energy. The
power-law index is referred to as $p_i$. Here, the subscript $i$ refers
either to electrons $i = e$  or to protons $i = p$. Energetic consideration
provides the constraint $\epsilon_B + \epsilon_e + \epsilon_p < 1$. 

The minimum Lorentz factors of the protons and electrons are readily obtained as
{
\begin{align}
    \gamma_{m, p} &\simeq 6 \hspace{0.3em} f_p \xi_p^{-1}E_{54}^{1/8} n_0^{-1/8} t_{3}^{-3/8}\epsilon_{p,-1},
    \label{Eq:gammap, min}\\
    \gamma_{m, e} &= 450 f_e\xi_e^{-1}E_{54}^{1/8} n_0^{-1/8} t_{3}^{-3/8}\epsilon_{e, -2}.
    \label{Eq:gammae, min}
\end{align}
}
where $f_i$ is a function of $p_i$ and is equal to $f_i = ({p_i-2})/({p_i-1})$ for $p_i>2$ and $f_i = \ln \left({\gamma_{m}}/{\gamma_{\max}}\right)$ for $p_i=2$ \citep{Sari_1998}. 

Another characteristic particle Lorentz factor is obtained by equating
the synchrotron cooling time to the dynamical time, providing the cooling
Lorentz factor of the particle. The synchrotron cooling time is given by $t_{syn} =  (6\pi m_i c)/(\gamma_i B^2 \sigma_{T,i})$, where $\sigma_{T,e}$ is the  Thomson cross-section and $\sigma_{T,p} = (m_e^2/m_p^2)\sigma_{T,e}$. This gives $\gamma_{c,i} = (6\pi m_i c)/(\sigma_{T,i}B^2\Gamma t)$, resulting in
{
\begin{align} \label{Eq:gamma, c}
\gamma_{c,e} &= 222.3~ t_{3}^{1/8} \epsilon_{B, -1}^{-1} E_{54}^{-3/8}n_0^{-5/8},
\\
\gamma_{c,p} &= 1.4 \times 10^{12}~ t_{3}^{1/8} \epsilon_{B,-1}^{-1} E_{54}^{-3/8}n_0^{-5/8}.
\end{align}
}
A proton synchrotron model requires the magnetic field to be large, and therefore, from Equations \eqref{Eq:gammae, min} and \eqref{Eq:gamma, c}, it follows that $\gamma_{c,e} < \gamma_{m,e}$, hence the electrons are in the fast 
cooling regime, and the electron distribution function is a broken power-law with index 
of 2 between $\gamma_{c,e}$ and $\gamma_{ m, e}$, and $p_e-1$ above $\gamma_{ m, e}$. However from Equation \eqref{Eq:gammap, min}, it is seen 
that $\gamma_{m,p} \ll \gamma_{c,p}$, meaning that the proton population 
is in the slow cooling regime.

The maximum Lorentz factor $\gamma_{\rm max,i}$ of the accelerated
particles is obtained by equating the acceleration time to
the synchrotron cooling time. It comes
\begin{align} 
\gamma_{\rm max,i} = ({6\pi q}/{\alpha \sigma_{T,i} B})^{1/2}
\label{eq:gamma_max}
\end{align}
where the numerical coefficient $\alpha$ prescribes the acceleration efficiency. The maximum Lorentz factor for protons is
{
\begin{align} \label{Eq:gamma, maxp} 
    \gamma_{\rm max,p} = 7.8 \times 10^{10} \hspace{0.3em} \alpha^{-1/2} E_{54}^{-1/16} n_0^{-3/16} t_{3}^{3/16} \epsilon_{B,-1}^{-1/4}.
\end{align}
}
Since $\gamma_{\rm max,p} < \gamma_{c,p}$, the proton distribution function is a single power-law above $\gamma_{m,p}$ with an exponential cutoff at the very high energy end of the proton distribution spectrum, producing an exponential cut-off in the resulting photon spectrum. For electrons, $\gamma_{\max,e}$ is $m_p/m_e$ times smaller than $\gamma_{\max,p}$, but as the electron-synchrotron flux at such high energies is very small, we omit the discussion on it here.

Each of those characteristic Lorentz factors in the particle
distribution functions is associated to a characteristic synchrotron frequency such that
$\nu = (3 q B \gamma^2 \Gamma)/(4\pi (1+z) m_i c)$. At these frequencies,
the observed synchrotron spectrum presents a spectral break. 
For the electrons, the observed spectrum is $F_\nu \propto \left(\nu^{1/3}, \nu^{-1/2}, \nu^{-p_e/2} \right)$ for $\left(\nu < \nu_{c,e}; ~\nu_{c,e} < \nu < \nu_{m,e}; ~ \nu_{m,e} < \nu \right)$, where $\nu$ is the observed frequency. The proton synchrotron spectrum is shaped as $F_\nu \propto \left(\nu^{1/3}, \nu^{-(p_p-1)/2} \right)$ for $\left( \nu < \nu_{m,p};~ \nu_{m,p} < \nu <
\nu_{\max,p} \right)$. 

The characteristic frequencies associated with particles at $\gamma_m$ are given by
{
\begin{align} 
 h\nu_{m,p} &=  7.6 \times 10^{-9}~f_p^2 E_{54}^{1/2} t_{3}^{-3/2} \epsilon_{B,-1}^{1/2}  \epsilon_{p,-1}^2 \xi_p^{-2}~\text{eV,} 
   \label{Eq:nu_m_p}\\
   h\nu_{m,e} &= 1.42 ~f_e^2 E_{54}^{1/2} t_{3}^{-3/2} \epsilon_{B,-1}^{1/2}  \epsilon_{e,-2}^2 \xi_e^{-2}~\text{eV.}
   \label{Eq:nu_m_e}
\end{align}
}
The cooling frequency for the electrons is
{
\begin{align}\label{Eq:nu_e,c}
    h\nu_{c,e} &= 0.35 ~E_{54}^{-1/2} t_{3}^{-1/2} \epsilon_{B,-1}^{-3/2} n_0^{-1}~\text{eV},
\end{align}
}
and the maximum frequency for the protons reads
\begin{equation} \label{Eq:vmax, p}
    h\nu_{\max, p} \sim 23~ \alpha^{-1} E_{54}^{1/8} n_{0}^{-1/8}t_{3}^{-3/8}\quad \text{TeV}. 
\end{equation}
Hence, the shock accelerated protons can emit synchrotron photons at energies as high as 10~ TeV above those detected by {LHAASO}.

The spectral flux is calculated as follows. 
The maximum power emitted by a single particle via the synchrotron process
at the observed peak frequency is $P_{\nu_{\max}} = ({2 m_i 
c^2 \sigma_{T,i} B \Gamma}/{9q}) (1+z)$. Assuming the total number of radiating particles to be $N_i = {4\pi 
\xi_i n r^3}/{3}$, the peak flux associated with the 
electron synchrotron process is
{
\begin{align} \label{Eq:Fpeak,e}
 F_{\nu_{\rm peak, e}} =\frac{N_e P_{\nu_{ e, \max}}}{4\pi d_L^2} = 2.7 \times 10^{-21}  \frac{\xi_e E_{54} \epsilon_{B,-1}^{1/2} n_0^{1/2}} {d_{L, 27}^{2}} \quad \text{erg~cm$^{-2}$~s$^{-1}$~Hz$^{-1}$},
\end{align}
}
where we normalised the luminosity distance $d_L$ to $10^{27}$~cm, given the proximity of GRB 221009A at  redshift $z = 0.151$ \citep{web:nasa:redshift, 2022GCN.32686....1C} corresponding to $d_L = 2.23 \times  10^{27}$~cm 
\citep[assuming a flat $\Lambda$CDM cosmology with $H_0$ = 69.6 km s$^{-1}$ Mpc$^{-1}$,
$\Omega_M= 0.286$, and $\Omega_\Lambda= 0.714$, ][]{2006PASP..118.1711W}. 
Since in our scenario the electrons are in the fast cooling  regime, analysing the spectrum reveals that $\nu_{c,e} < \nu_{o}< \nu_{m,e} < \nu_{XRT}$, where $\nu_{XRT} = 0.3$~keV is the low energy threshold of the XRT instrument and $\nu_o \sim 1$~eV is the typical frequency of the optical band. 
The flux at frequencies lower than $\nu_{m,e}$ and greater than $\nu_{c,e}$ is given by
\begin{align}
\label{Eq:flux_e_1}
F_{\nu, e}  = F_{\nu_{\rm peak, e}} (\nu/
\nu_{c,e})^{-{1/2}},
\end{align}
while the flux at frequencies higher than $\nu_{m,e}$ is given by
\begin{align}
\label{Eq:flux_e_2}
F_{\nu, e}  = F_{\nu_{\rm peak, e}} (\nu_{m,e}/
\nu_{c,e})^{-{1/2}} (\nu/\nu_{m,e})^{-{p_e/2}}.
\end{align}

Similarly, the peak flux associated with the proton-synchrotron process is given by {
\begin{align}\label{Eq:Fpeak, p}
     F_{\nu_{\rm peak, p}} =\frac{N_p P_{\nu_{p, \max }}}{4\pi d_L^2}= 1.46 \times 10^{-24} \frac{\xi_p E_{54} \epsilon_{B,-1}^{1/2} n_0^{1/2}} {d_{L, 27}^{2}}\quad \text{erg~cm$^{-2}$~s$^{-1}$~Hz$^{-1}$},
\end{align} 
}
and the spectrum within the frequency range $\nu_{m,p} \leq \nu \leq \nu_{\max,p}$ for the slow-cooling synchrotron process is a power-law 
\begin{align}
\label{Eq:flux_p}
F_{\nu, p} = F_{\nu_{\rm peak, p}} (\nu/\nu_{m,p})^\frac{-(p_p-1)}{2}.
\end{align}

\section{Limitations on the model parameters: An Analytical approach}

\label{sec:Limits}

The best fit to the temporal decay of the X-ray flux as observed by the \textit{swift-}XRT consists of five breaks at times between 
$T_0 + 3.27 \times 10^4$~s to $T_0 + 4.5 \times 10^6$~s. 
The corresponding decay indices at the break times are presented in the XRT catalogue of the burst \citep{2009MNRAS.397.1177E}. 
At observed times $t < 3\times 10^4$~s, the decay index in the XRT band is identified to be $\sim -3/2$.\footnote{\url{https://www.swift.ac.uk/xrt_live_cat/01126853/}} At later times, the lightcurve becomes steeper, and may be associated with a jet break. In order to reproduce the temporal decay in the XRT band with $\nu_c < \nu_m < \nu_{XRT}$ 
(fast cooling regime, as required by the {electron} synchrotron model), the condition $p_e \sim 8/3 \sim 2.67$ must be satisfied as $F_\nu \propto t^{\frac{2-3p}{4}}$.\footnote{In the slow cooling regime, the requirement on the electrons power law index is even higher with $p \sim 3$ for $\nu_m < \nu_{XRT} < \nu_c$.}

On the other hand, the TeV temporal decay is $-1.115 \pm 0.012$ \citep{doi:10.1126/science.adg9328}. 
This result is challenging for the SSC model, as it seem outside the expected temporal slope for electrons with a power law index of $2.67$, which is $-1.625$ \citep{2001ApJ...548..787S} in the fast cooling regime.\footnote{In the slow  cooling regime, assuming  $\nu_m < \nu_{XRT} < \nu_c$, the expected temporal decay in the TeV band is $-2$, for $p=3.0$.}
We can therefore exploit the  proton-synchrotron model, with proton power law index of $\approx 2.2- 2.3$ in the slow cooling regime, as explained above. This is the index needed to explain the {LHAASO} data. In fact, setting $p_e \equiv p_p \sim 8/3$ leads to an energy crisis, as the energy budget in the proton-synchrotron component will be very high. Therefore, in presenting our results for GRB 221009A, we will set a proton power law index $p_p = 2.3$, but allow for a different value of $p_p = 2.2$ in interpreting the TeV data. We will keep the constraint $ p_e = 8/3$ in order to satisfy the XRT temporal decay.

\subsection{Constraints for a proton-synchrotron model}
\label{sec:constraints_ps}
\begin{figure}[t!]
    \centering
     \begin{tabular}{cc}
    \includegraphics[width=0.45 \textwidth]{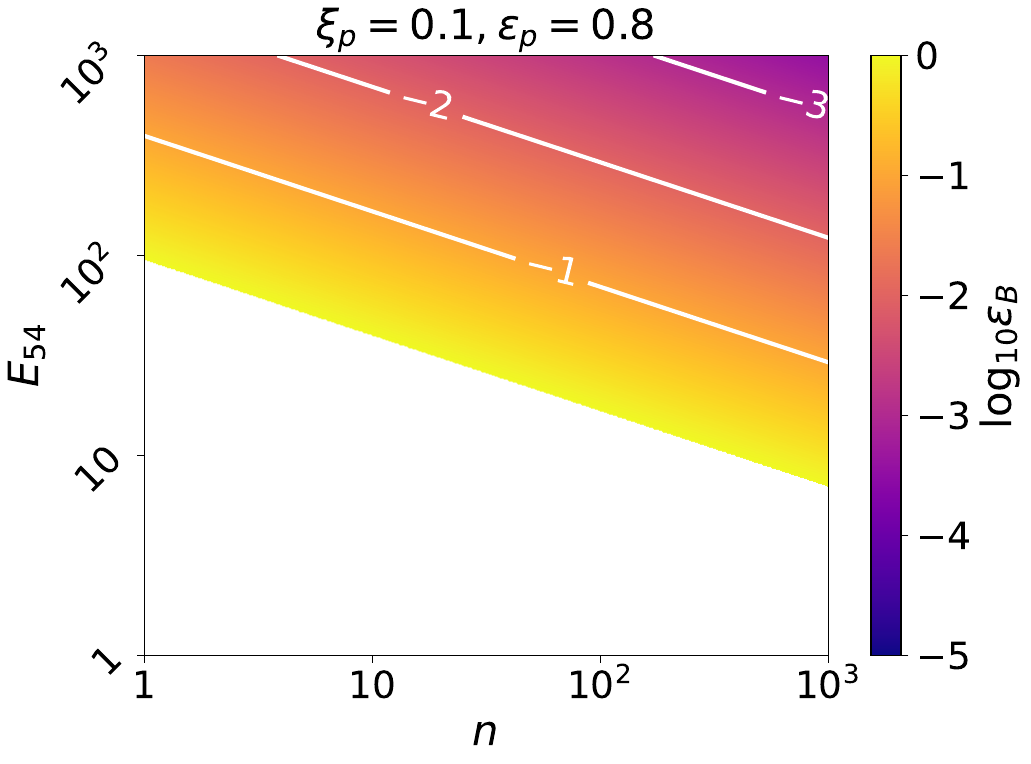}   &
        \includegraphics[width=0.45 \textwidth]{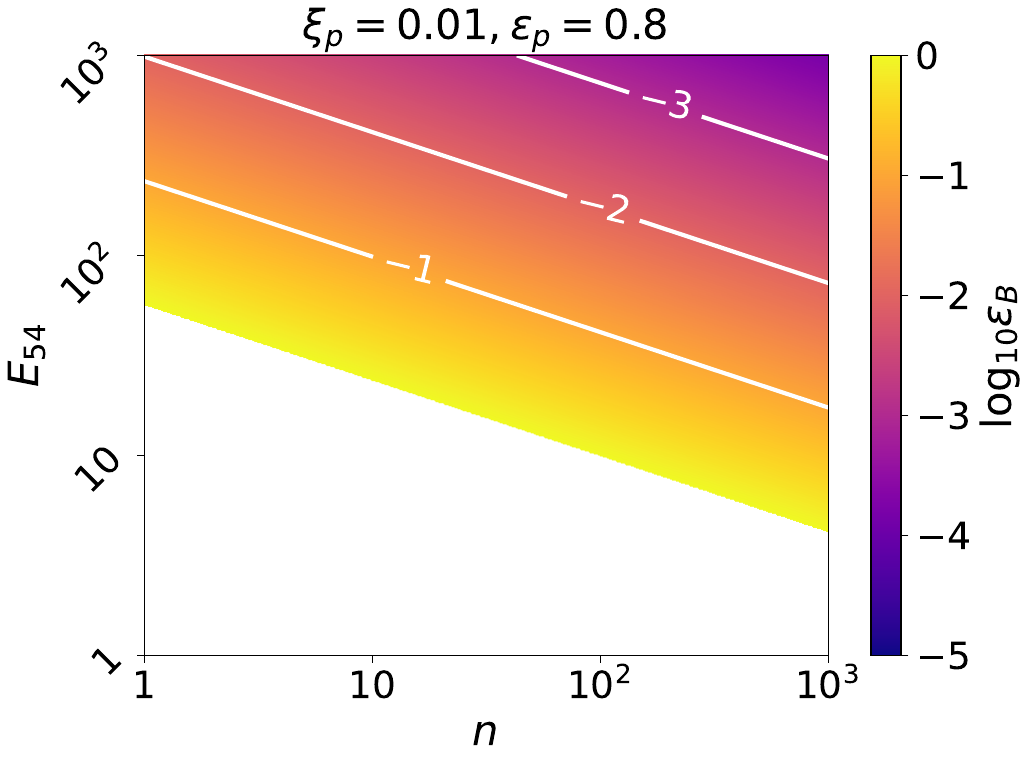}  \\
        \includegraphics[width=0.45 \textwidth]{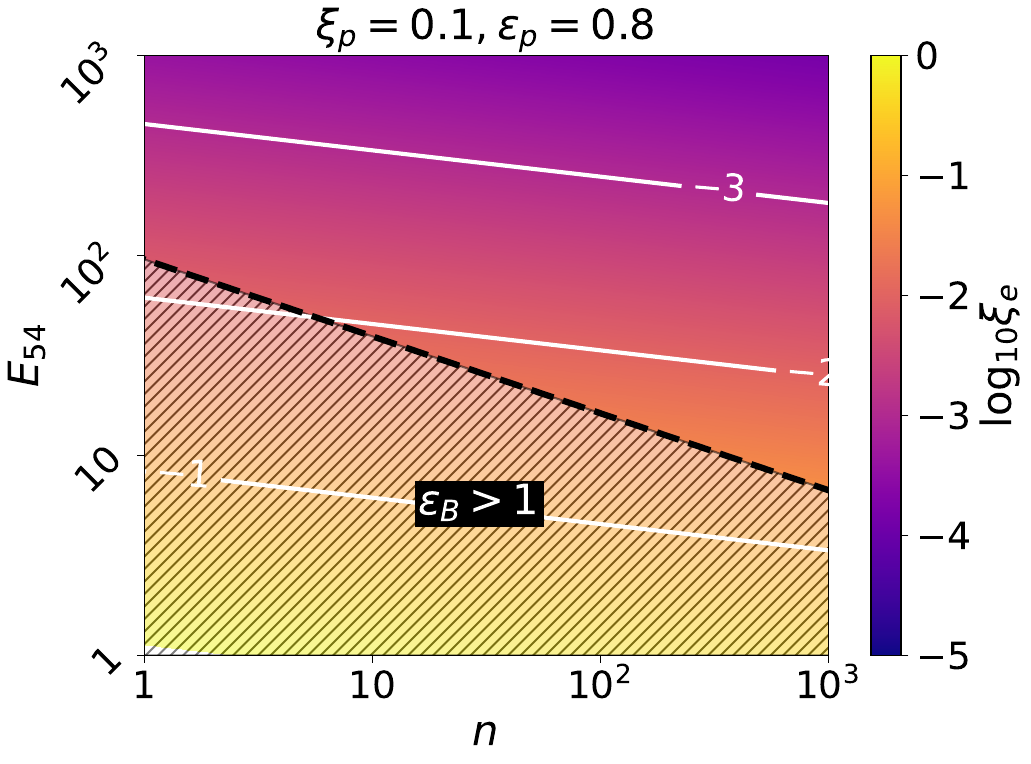}&
       \includegraphics[width=0.45 \textwidth]{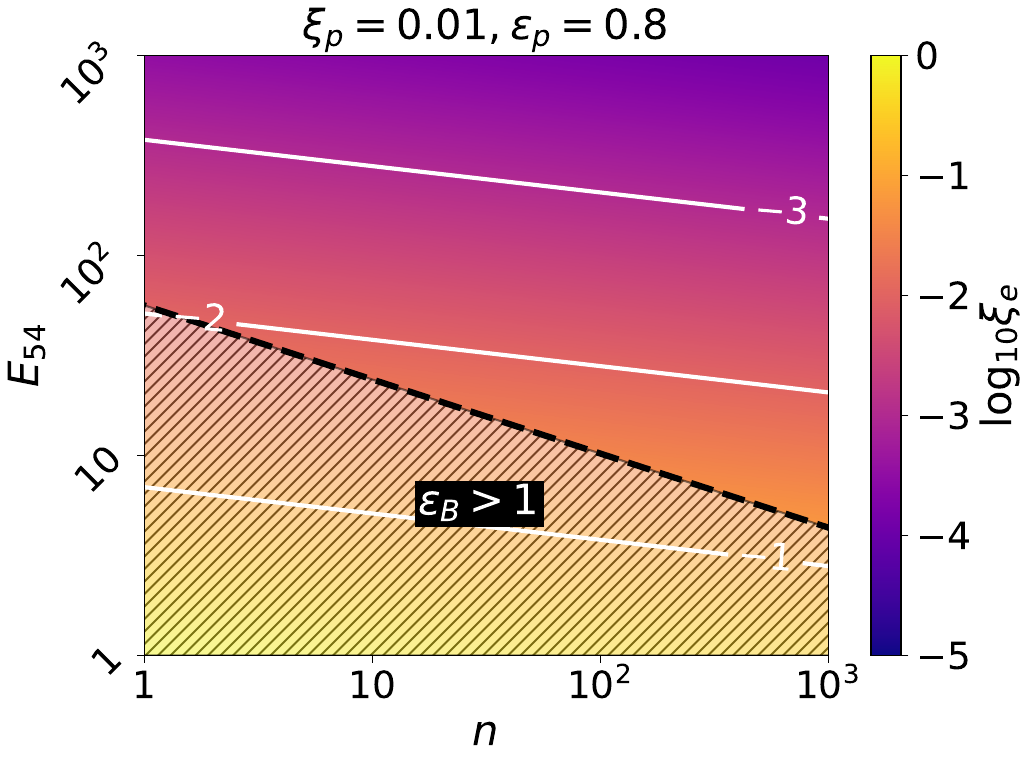}\\
       \multicolumn{2}{c}{\includegraphics[width=0.45 \textwidth]{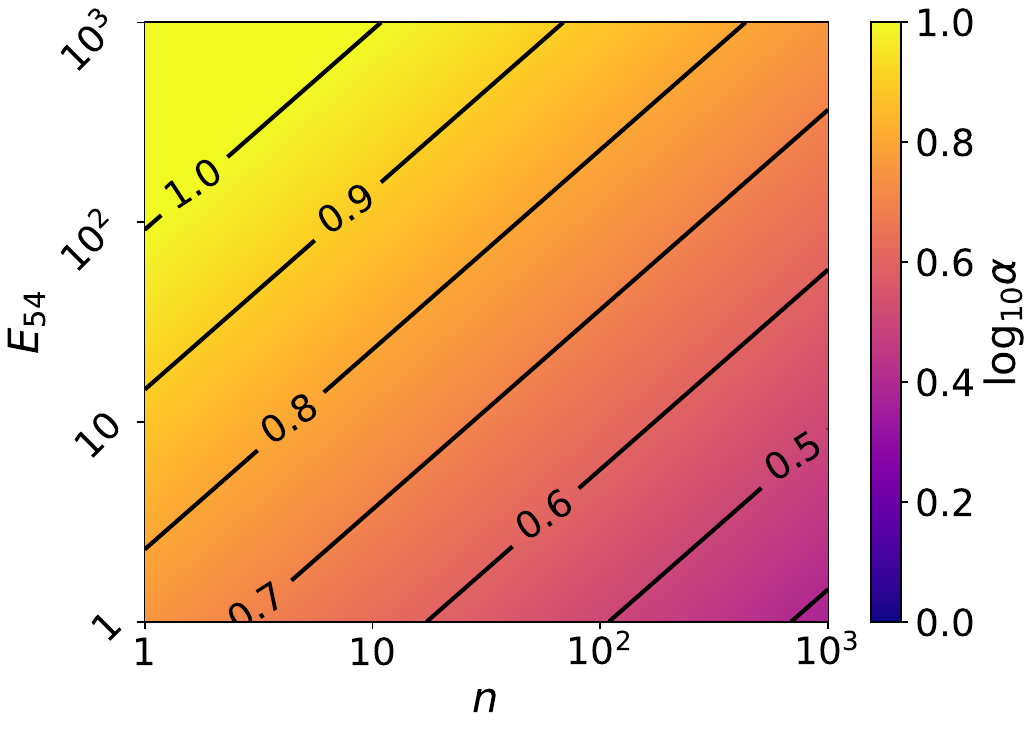}}

    \end{tabular}
    \caption{
    Parameters $\epsilon_B$ (top) and $\xi_e$  (middle)  as functions of the total energy and ambient density, $E_{54}$ and $n_0$ for $\epsilon_p = 0.8$  where assumed fraction of 10\% of the protons are accelerated into a power law  ($\xi_p = 0.1$, left) and 1\% accelerated into a power law
    ($\xi_p = 0.01$, right). Bottom: the acceleration efficiency, $\alpha$ as a function of $E_{54}$ and $n_0$. 
    The plots are obtained from Equations \eqref{Eq:e_B}, \eqref{Eq:xi_e},  and \eqref{Eq:alpha}, respectively. The white region is forbidden as it corresponds to  $\epsilon_B > 1$ as seen from the top panel.
    The assumptions $p_e =8/3$ and $p_p = 2.3$ are enforced for these figures. 
    }
   \label{Fig:constraints} 
\end{figure} 
We use the observed flux of GRB 221009A to constrain the values
of the free model parameters. We start with the
flux at 1 TeV as observed by the LHAASO experiment at time $t=235$~s after the trigger. By equating the LHAASO reported flux
$[\nu F_\nu]_p \approx 2  \times 10^{-6}$~ergs~cm$^{-2}$~s$^{-1}$  \citep{doi:10.1126/science.adg9328} and the 
expected proton-synchrotron flux given in Equation \eqref{Eq:flux_p} at 1 TeV, the fractional energy of the magnetic field as a 
function of the other free parameters is,
{
\begin{align}
\label{Eq:e_B}
\epsilon_{B,-1} =&~ 0.1 \cdot {{(8.94\cdot {{10}^{-8}})}^{\frac{4}{p_p+1}}} {{(2.53\cdot{{10}^{18}})}^{\frac{2 \left( p_p-1\right) }{p_p+1}}}\, {{f_p}_p}^{-\frac{4 {p_p}-4}{{p_p}+1}} \xi_p^{\frac{4(p_p-2)}{(p_p+1)}}E_{54}^{-\frac{p_p+3}{p_p+1}}\epsilon_{p,-1}^{-\frac{4(p_p-1)}{(p_p+1)}}n_0^{-\frac{2}{p_p+1}},  \nonumber 
\end{align}
}
which upon setting $p_p = 2.3$ simplifies to
{
\begin{align}
\epsilon_{B,-1} =&~9.1 \times 10^{5}~ \xi_p^{4/11} E_{54}^{-53/33}\epsilon_{p,-1}^{-52/33} n_0^{-20/33}.
\end{align}
}
This value of $\epsilon_{B}$ may seem large, but so should be the kinetic energy of this 
burst, leading in fact to values of $\epsilon_B$ smaller than unity, see the top panel plot on Figure \ref{Fig:constraints}.

Similarly, the acceleration efficiency parameter $\alpha$ is obtained by comparing
$\nu_{\max,p}$ from Equation \eqref{Eq:vmax, p} with the observed energy of 7 TeV photon at $t = 235$~s, 
\begin{align}\label{Eq:alpha}
\alpha = 5.7~E_{54}^{1/8}n_0^{-1/8}.
\end{align}

To obtain the values of the other model parameters, we refer to \textit{swift}-XRT data available at much later times, 
where XRT data is available and the emission is clearly in the "afterglow" phase in the \textit{swift}-XRT band as well. 
We use Equation \eqref{Eq:flux_e_2}, which gives the expected energy flux above $\nu_{m,e}$, together with the observed \textit{swift}-XRT flux at 3300~s of $1.087\times10^{-7} \rm erg~ \rm s^{-1} \rm cm^{-2}$ (at 0.3 keV) and include Equation \eqref{Eq:e_B} to get the injection fraction of electrons,
{
\begin{align}
\xi_e \simeq &~  {({8.94 \cdot {{10}^{-8}})}^{\frac{1}{{p_p}+1}}}\, {({2.53 \cdot {{10}^{18}})}^{\frac{{p_p}-1}{2 {p_p}+2}}}\,  {{0.025}^{-\frac{1}{{p_e}-2}}} {{6.44}^{-\frac{{p_e}}{2 \left( {p_e}-2\right) }}}\,  10^{\frac{1-p_e}{{p_e}-2}}\,{{{{{f_p}}_e}}^{\frac{{p_e}-1}{{p_e}-2}}}\, {{{{{f_p}}_p}}^{\frac{1-{p_p}}{{p_p}+1}}}\, \nonumber\\
&~~~ \times {{{E_{54}}}^{\frac{2 {p_p}-{p_e}+4}{\left( 2 {p_e}-4\right)  {p_p}+2 {p_e}-4}}}\, {{{{{\epsilon}}_{e,-2}}}^{\frac{{p_e}-1}{{p_e}-2}}}\, {{{{{\epsilon}}_{p,-1}}}^{\frac{1-{p_p}}{{p_p}+1}}}  {{{{\xi}_p}}^{\frac{{p_p}-2}{{p_p}+1}}}\, {{{n_0}}^{-\frac{1}{2 {p_p}+2}}}\,.\nonumber
\end{align}
}
Setting $p_e = 8/3$ and $p_p = 23/10$ gives
{
\begin{align}
\label{Eq:xi}
\xi_e\simeq&~ 0.11~ E_{54}^{89/66} \xi_p^{1/11} \epsilon_{e,-2}^{5/2} n_0^{-5/33} \epsilon_{p,-1}^{-13/33}.
\end{align} 
}
By balancing the \textit{swift-}UVOT band flux of $1.22 \times 10^{-8}~ \rm erg~ \rm s^{-1} \rm cm^{-2}$~at observed energy 4.77~eV and observed time $t = 4000$~s 
with the predicted synchrotron flux from Equation \eqref{Eq:flux_e_1} and including Equations \eqref{Eq:e_B} and \eqref{Eq:xi}, we get
{
\begin{align}
\epsilon_{e,-2} \approx&~10 \cdot{{0.025}^{\frac{1}{{p_e}-1}}} {{0.06}^{\frac{{p_e}-2}{{p_e}-1}}} {{6.44}^{\frac{{p_e}}{2 \left( {p_e}-1\right) }}}{E_{54}}^{-1} {{\mathit{f_p}}_e}^{-1}, \nonumber
\end{align}
}
which gives for $p_e = 8/3$
{
\begin{align}
\label{Eq:e_e}
\epsilon_{e,-2} \approx&~  3.9~E_{54}^{-1}.
\end{align}
}
This enables to write the parameter $\xi_e$ as
\begin{align}
   \xi_e &= 0.06 \cdot {({8.94 \cdot {{10}^{-8}})}^{\frac{1}{{p_p}+1}}}\, {({2.53 \cdot{{10}^{18}})}^{\frac{{p_p}-1}{2 {p_p}+2}}}\, {{{{{f_p}}_p}}^{\frac{1-{p_p}}{{p_p}+1}}}\,{{{E_{54}}}^{\frac{-2 {p_p}-3}{2 {p_p}+2}}}\ {{{{{\epsilon}}_{p,-1}}}^{\frac{1-{p_p}}{{p_p}+1}}}\, {{{{\xi}_p}}^{\frac{{p_p}-2}{{p_p}+1}}\, {{{n_0}}^{-\frac{1}{2 {p_p}+2}}}}\,, \nonumber
\end{align}
which simplifies to
\begin{align}
\label{Eq:xi_e}
   \xi_e &= 3.2~\xi_p^{1/11} E_{54}^{-38/33} n_0^{-5/33}\epsilon_{p,-1}^{-13/33},
\end{align}
for our chosen index values.

The values of $\epsilon_B$, $\alpha$ and  $\xi_e$ , as constrained by Equations \eqref{Eq:e_B}, \eqref{Eq:alpha} and \eqref{Eq:xi_e} are plotted in Figure \ref{Fig:constraints} as functions of $E_{54}$ and $n_0$ for $\epsilon_p = 0.8$ and two choices of $\xi_p $, namely $\xi_p = 0.1$ (left column) and $\xi_p  = 0.01$ (right column). 
Satisfying $\epsilon_B < 1$ directly requires that $E_{54}$, $n_0$ and $\epsilon_p$ should be large while $\xi_p$ needs to be small.
Overall, we could constrain the parameters $\epsilon_B$, $\epsilon_e$, $\xi_e$ and $\alpha$ as functions of the other free model parameters.
To provide satisfactory constraints with
$\epsilon_e + \epsilon_B + \epsilon_p < 1$, the kinetic energy of this 
burst must be large with $E_{54} > 10$. Yet this is not too large compared to the 
prompt total isotropic energy. In fact, this high kinetic energy would correspond to an efficiency of around 10\%, typical of other GRBs \citep[see e.g.][]{2007ApJ...655..989Z, 2016MNRAS.461...51B}. We note that our goal here is to provide a set of parameters that could potentially explain the TeV observations via the proton synchrotron process and not to determine the best possible parameter values. 

We thus find that a requirement of our model is that accelerated protons contribute for
most of the internal energy of the shock, with $\epsilon_p \gtrsim 0.1$. The magnetic field needs to be strong, 
$\epsilon_B \gtrsim 10^{-2}$, and the circumburst density should be high, $n_0 \gtrsim 10$.  All these
require a high kinetic energy. Furthermore, the model requires that only a relatively small
fraction of electrons and protons achieve a power-law distribution
behind the shock, \textit{i.e.} $\xi_e \sim 10^{-2}$ and $\xi_p \lesssim 10^{-1}$, and that their spectral indices be different, {$p_e \neq p_p$}. 

\begin{figure}[ht!]
     \centering
     \includegraphics[width=0.4\textwidth]{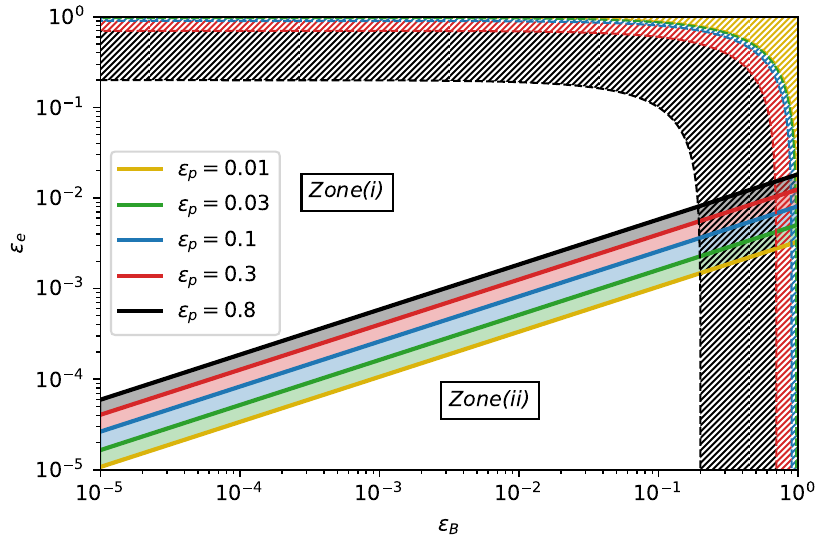}\\ 
    \caption{ {The parameter regime $\epsilon_B$, $\epsilon_e$ and $\epsilon_p$ is explored in the context of $p_e = 8/3, p_p = 23/10, E_{54} = 50$, $n_0 = 50$, corresponding to $\xi_e = 0.01$ and $\xi_p = 0.04$ for the condition \eqref{Eq:PSynch/Ic cond}. Notably, the proton-synchrotron process dominates mostly in regions where $\epsilon_B > \epsilon_e$ (Zone (ii)), while the SSC process takes over when $\epsilon_e > \epsilon_B$ (Zone (i)). Zone (ii) is shown to have different boundaries based on the value of $\epsilon_p$ and these boundaries visually distinguish both the regions. It is clear that the dominance of proton synchrotron emission becomes more prominent as $\epsilon_p$ increases. The hatched regions represent the condition $\epsilon_B+\epsilon_p+\epsilon_e \leq 1$ for each $\epsilon_p$ value and regarded as forbidden zones.}}  
    \label{Fig:PSynch/Ic}
\end{figure}
\subsection{Constraints imposed by the synchrotron-self Compton (SSC) emission.}
\label{sec:consts_IC_thry}

In the SSC emission mechanism, the electrons that emit the low-energy 
synchrotron-photons inverse Compton (IC) scatter the same photons to higher energies, thus contributing to the high energy component of the afterglow spectrum. As we derive in details in appendix  \ref{app:IC}, for the parameters chosen, this component is sub-dominant. Here  we present the results only for $p_e = 8/3$ and $p_p = 23/10$, but the general trend applies for other values of the injection index $p_p$.
Within the framework of our model, the Klein-Nishina effect for the IC component can be neglected. This is shown by using Equations \eqref{Eq:Gamma}, \eqref{Eq:gammae, min} and \eqref{Eq:nu_m_e} at $t = 235$~s, to find 
\begin{align} \label{Eq:KNC}
     \frac{\gamma_{m,e} h\nu_{m,e}}{\Gamma} &\sim 0.15~ E_{54}^{5/33} \epsilon_{p,-1}^{13/33} n_{0}^{5/33} \xi_p^{-1/11} ~\rm MeV. 
\end{align}
This result is in the order of (and even smaller than) $m_e c^2$, the energy
at which the Klein-Nishina effect becomes important. One therefore only expects small
modifications, if any, around the peak of the IC component.

The characteristic frequencies of the IC  spectral component
for this burst are presented in  Appendix \ref{app:IC}. Equation \eqref{Eq:Ic_min_fqcy} gives the 
observed peak energy of the IC spectrum to be
around $h\nu_{m, IC} \sim 30$~GeV.
{
To determine the criterion that governs the dominance of the proton-synchrotron component over the SSC component, we compare the fluxes at 1 TeV and at 235 s as $[\nu F_\nu]_{p}/[\nu F_\nu]_{IC} \gtrsim 1 $ \citep{2001ApJ...559..110Z}. We can expand it using Equations \eqref{Eq:flux_p}, \eqref{Eq:Fpeak, p}, \eqref{Eq:nu_m_p} and \eqref{Eq:flux_IC} to get the following:
\begin{align} \label{Eq:PSynch/Ic cond}
    0.15~E_{54}^{3/40} n_0^{7/12}\xi_e^{4/3}\xi_p^{-3/10} \epsilon_{B,-1}^{199/120} \epsilon_{p,-1}^{13/10} \epsilon_{e,-2}^{-10/3} &\gtrsim 1.
\end{align}
}
{This condition is displayed in Figure \ref{Fig:PSynch/Ic}. The findings depicted in Figure \ref{Fig:PSynch/Ic} indicate the notion that $\epsilon_B$ must exceed $\epsilon_e$ in order to satisfy the above condition.  Also it is evident that a near equipartition value of $\epsilon_p$ reinforces the significant contribution from proton -synchrotron process. 
}

From Equation \eqref{Eq:flux_IC}, the flux at 1 TeV and $t_3 = 0.235$, using the results of Equations \eqref{Eq:e_B} and \eqref{Eq:xi} reads
\begin{align}\label{Eq:flux_IC_mod} 
[\nu F_\nu]_{IC}|_{1 {\rm ~TeV}} &= 3.4 \times 10^{-14}~ E_{54}^{313/396} \epsilon_{p,-1}^{182/99} n_0^{247/396} \xi_p^{-42/99} ~~\text{ergs~cm$^{-2}$~s$^{-1}$},
\end{align}

where we assumed that the TeV band is above the frequency of the IC spectral peak, and neglected Klein-Nishina effects. If anything, this effect would further reduce the observed flux in the TeV band, and allow for a larger parameter space in which the proton synchrotron mechanism dominates.
This value is $\approx 8$ orders of magnitude lower than the observed flux at the TeV band, implying that the electron SSC mechanism is sub-dominant at these energies for the constraints we derived.

\section{Lightcurve and spectra}
\label{sec:Results}

\begin{figure}[ht!]
    \centering
     \includegraphics[width=0.6\textwidth]{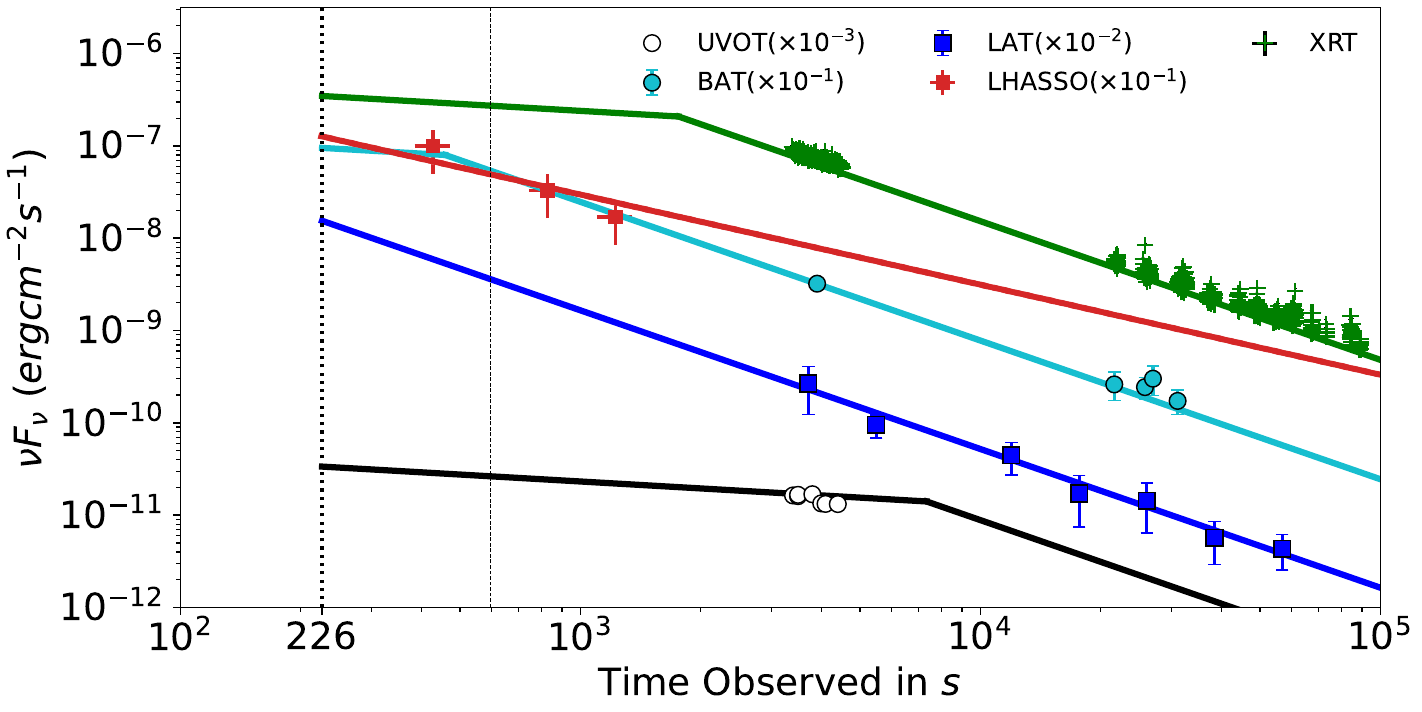}\\ 
    \caption{The multi-wavelength afterglow light curve for the
    proton synchrotron model along with the
    observed data of GRB 221009A for $p_e = 8/3, p_p =23/10, E_{54} = 50$,
     $n_0 = 50 $, $\epsilon_{p,-1} = 8$  and $\xi_p = 0.04$, corresponding to $\epsilon_{B,0} = 0.186$, $\epsilon_{e,0} = 0.0014$ and $\xi_e = 0.01$.  The \textit{swift}- UVOT (white filter) and BAT data were retrieved from Tables 4 and 6 of \citet{2023ApJ...946L..24W} and the \textit{fermi}-LAT data were obtained from Table 5 of \citet{2023ApJ...946L..23L}. The optical data is corrected for galactic extinction with $A_{\rm v} = 5.4$ \citep{2023ApJ...946L..25S, 2023ApJ...946L..22F}. The black dotted line at 226~s marks the onset of the afterglow phase corresponding to \citet{doi:10.1126/science.adg9328}, while the black dashed line at $T_0 = 597$~s marks the afterglow onset determined by \citet{2023arXiv230314172L}.}  
    \label{Fig:Light_curve}
\end{figure}

 The observed lightcurve of the afterglow of GRB 221009A is shown
in Figure \ref{Fig:Light_curve} alongside the electron and proton synchrotron 
components estimated from our model. The two different set of parameters used in this section are summarized in Table \ref{tab:1_table}. We set the parameters to $p_e = 8/3$,  $p_p = 2.3$, $ E_{54} = 50, n_0 = 50$, $\epsilon_{p,-1} = 8$ and $\xi_p = 0.04$, resulting in $\epsilon_{B,0} = 0.186$, $\epsilon_{e,0} = 0.0014$ and $\xi_e = 0.01$. 
The onset of the afterglow phase at 226~s after the burst trigger is marked by the dotted line \citep{doi:10.1126/science.adg9328} whereas the dashed line drawn at t = 597~s marks the end of the prompt duration estimated by \citet{2023arXiv230314172L}. 
Overall, this figure demonstrates that the
proton synchrotron model presented here is capable of explaining the observed temporal features of the afterglow of GRB 221009A in several different energy bands.

\begin{figure}
    \centering
    \begin{tabular}{cc}
    \includegraphics[width=0.4\textwidth]{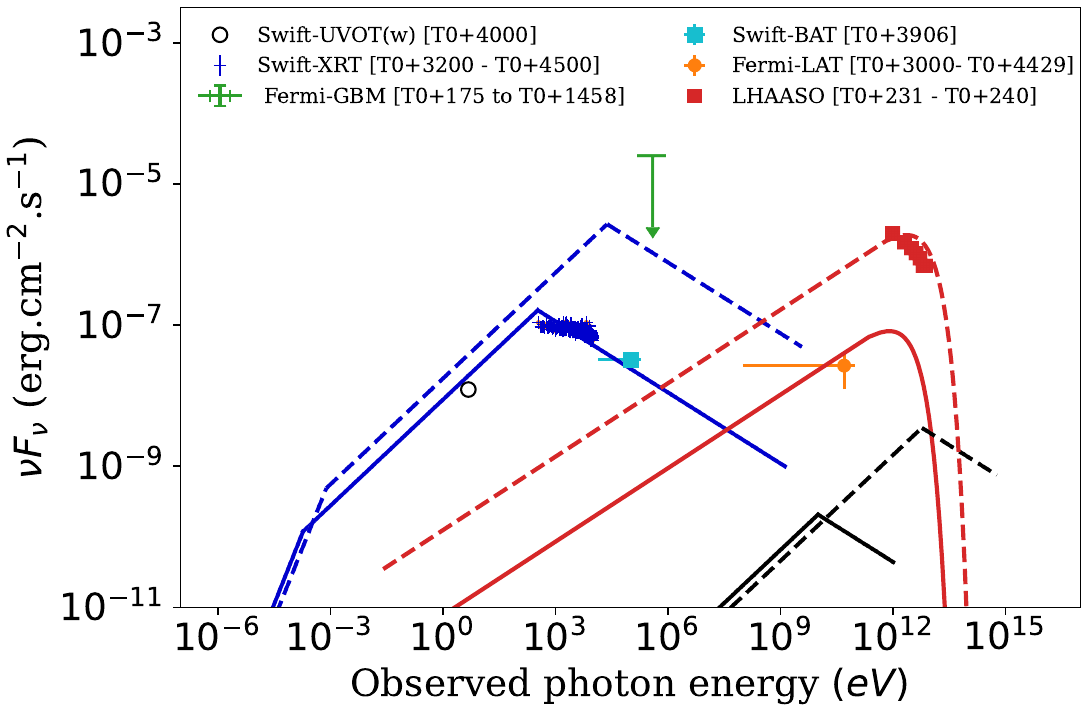}&
    \includegraphics[width=0.4\textwidth]{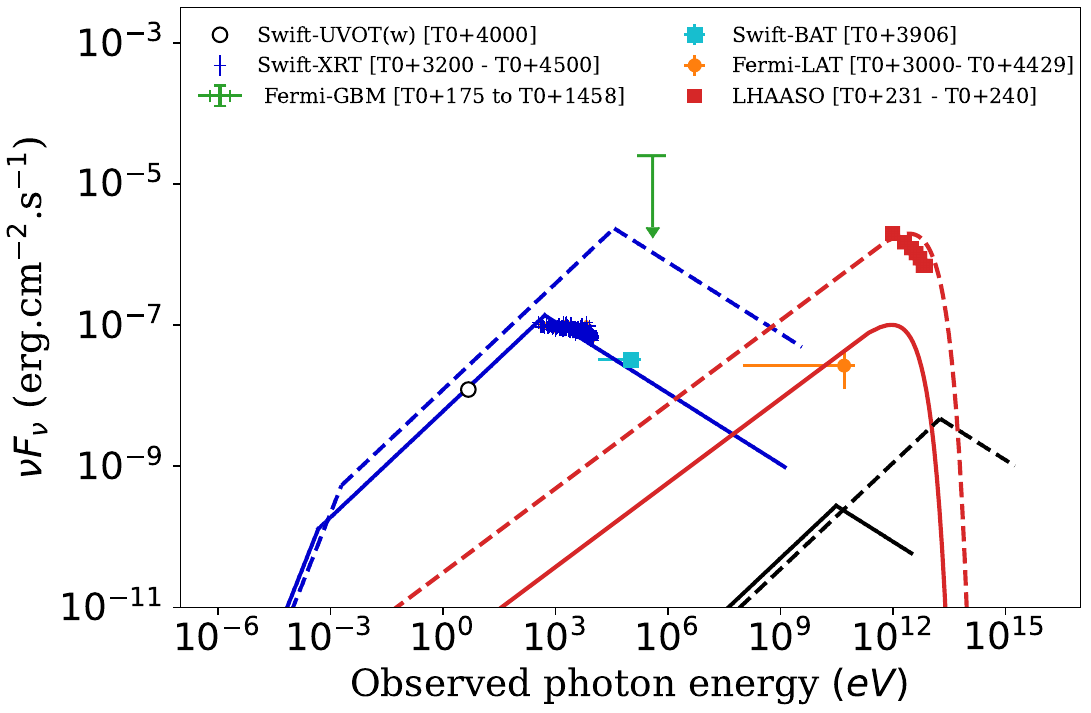} 
    \\
    \end{tabular}
    \caption{ Left: the spectral energy distributions for $p_e = 8/3$ and
    $p_p = 23/10$ at different times, $t = 235$~s (dashed) and $t = 4000$~s (solid). The
    parameters  are $E_{54} = 50$, $n_0 = 50 $, $\epsilon_{p,-1} = 8$  and
    $\xi_p = 0.04$, leading to {$\epsilon_{B} = 0.186$, $\epsilon_{e} = 0.0014$,}  $\xi_e = 0.01$ and  $\alpha = 6$.
     Right: the spectral energy distributions for
     for $p_e = 8/3$ and  $p_p = 22/10$, the other parameters are $E_{54} = 30$,
    $n_0 = 30 $, $\epsilon_{p,-1} = 8$  and $\xi_p = 0.1$, corresponding to {$\epsilon_{B} = 0.17$,
    $\epsilon_{e} = 0.002$,}  $\xi_e = 0.01$ and  $\alpha = 6$.
    The blue line represents the electron synchrotron component, the red line represents the proton synchrotron emission and the black line specifies the IC component.}  
    \label{Fig:SED_plot_p_2.3.2}
\end{figure}

We further produce the afterglow spectrum at two
different times, namely at $t = 235$~s and $t = 4000$~s,
and display it in Figure \ref{Fig:SED_plot_p_2.3.2} (both in the left and right panels). The
spectral energy distributions (hereinafter SEDs) consist of three main components. They are produced by the
electron-synchrotron, the proton-synchrotron and the IC processes and are
respectively shown by the blue, red and black lines. Inspection of Figure
\ref{Fig:SED_plot_p_2.3.2} (left), obtained for $p_e = 8/3$ and
$p_p=2.3$, shows that the SED of the electron-synchrotron component at
4000~s satisfies the observed data in the optical and X-ray bands as
designed in our analytic approach. At the same time, the
proton synchrotron component accounts for the LHAASO flux, also
marginally satisfies the LAT flux. 

We then search for solutions with a smaller proton
index, $p_p = 2.2$. 
This is advantageous as it results in a lower
burst energy and external
density. The right panel of Figure \ref{Fig:SED_plot_p_2.3.2} shows the SEDs
for the parameters $E_{54} = 30$, $n_0 = 30 $,  $\xi_p = 0.1$, $\epsilon_{p,-1} = 8$, resulting in $\epsilon_{B} = 0.17$,  $\xi_e = 0.01$,  
$\epsilon_{e} = 0.002$ and $\alpha = 6$. In addition,
under this assumption, the required total burst energy $E$ is lower than for $p_p = 2.3$. Hence the prompt radiative efficiency is increased here.   
We see that, in this case too, the LAT flux at 4000~s can be marginally explained
by the proton synchrotron process.

The close to equipartition values of $\epsilon_B$ and  $\epsilon_p$ associated with all the SEDs are
as anticipated for the proton-synchrotron model, see Equations \eqref{Eq:e_B}
and \eqref{Eq:xi_e}. Indeed, protons are more massive
than electrons and to radiate a substantial amount of energy, they
need a strong magnetic field. We further find that the IC components
(black lines in the SEDs shown in Figure \ref{Fig:SED_plot_p_2.3.2})
are subdominant at all time bins in both 
scenarios considered, since the magnetic field energy density is large compared to the
electron energy density \citep[$U_B \gg U_e$, see e.g.][]{1986rpa..book.....R}. Indeed, for the constraints we derived, using $p_e = 8/3$ and $p_p = 23/10$, one obtains  
\begin{align*}
\begin{array}{lll}
    &U_B  &= 2.1  \times 10^6 ~  E_{54}^{-179/132} \epsilon_{p,-1}^{-52/33} n_0 ^{19/132} t_3 ^{-3/4} \xi_p^{4/11},\\
\text{and} &   U_e  &= 0.14~{{{E_{54}}}^{-\frac{7}{8}}}\, {{{n_0}}^{\frac{7}{8}}}\, {{{t_3}}^{-\frac{3}{8}}}.
\end{array}
\end{align*}
The corresponding ratios of the energy densities in our models are given in Table \ref{tab:1_table}. 

The lightcurve of the TeV emission is shown to have a break followed by a
steeper temporal decay at time $T_0+896~(+230, -110)$~s 
\citep{doi:10.1126/science.adg9328}. This break could be obtained by
the crossing of the maximum synchrotron frequency $\nu_{\rm max}$ through
the {LHAASO} energy band. For the parameters we derived in case
$p_e = 8/3$ and $p_p =2.3$, the time at which $\nu_{\rm max}$ equals
1 TeV is $t_3 = 42$. In principle, it is possible to use this property to
better constrain the acceleration efficiency, $\alpha$ and the other model parameters. However,
this  constraint depends on the exact parametrization of the proton
distribution function at the highest energies, and therefore would
bring only little insight into the model, apart from better constraining $\alpha$. 

\begin{table}[b]
\centering
\caption{Parameters used to construct the SEDs in Figure \ref{Fig:SED_plot_p_2.3.2}.}
\label{tab:1_table}
\centering
\resizebox{\columnwidth}{!}{
\begin{tabular}{|l||cccccc||cccc||c||c|}
\hline
& $p_e$ & $p_p$ & $E$ & $n$ &  $\xi_p$ & $\epsilon_p$ & $\epsilon_B$ & $\epsilon_e$& $\xi_e$  & $\alpha$ & $\eta$ & $U_B/U_e$ \\
\hline
Figure \ref{Fig:SED_plot_p_2.3.2} left & 8/3 & 2.3 & $5 \times 10^{55}$~erg & 50~cm$^{-3}$ & 0.04 & 0.8 & 18.57$\times 10^{-2}$ & 1.4 $\times 10^{-3}$ & 0.01 & 6 & $5.7\%$ & $8.1 \times 10^3~{{t_3}}^{-3/8}$ \\
Figure \ref{Fig:SED_plot_p_2.3.2} right & 8/3 & 2.2 & $3 \times10^{55}$~erg & 30~cm$^{-3}$ & 0.1 & 0.8 &  17 $\times 10^{-2}$ & 2 $\times 10^{-3}$ & 0.01 & 6 & $9.1\%$ & $5.2 \times 10^3~{{t_3}}^{-3/8}$ \\
\hline
\end{tabular}
    }
\end{table}

\section{Discussion} \label{sec:discussion}
\subsection{Comparison between GRB 221009A and GRB 190114C}

Similar to GRB 221009A, 
GRB 190114C is another long GRB with a VHE afterglow emission
observed in the band between 0.2 and 1
TeV  \citep{MAGIC:2019irs}. This GRB has an isotropic equivalent energy
$E_{iso} \simeq 2.5 \times 10^{53}$~erg. The redshifts of both GRB 190114C
($z=0.4245$) and GRB 221009A are low, $z<0.5$. As a result, the  detectability
of $\geq 1$~TeV photons, if produced in the source, is high, since they
are only weakly EBL attenuated \citep[see e.g.][]{2021Univ....7..146F}. Similarly to
GRB 221009A, a complete set of multi-wavelength observational data is available
for GRB 190114C afterglow \citep[see \textit{e.g.}][for a summary of those
observations]{2019Natur.575..455M}. In  \citet{2022arXiv221002363I}, we
considered the proton synchrotron mechanism to explain the VHE afterglow of GRB 190114C.

One can therefore compare the parameters we derived for GRB 221009A to those we obtained for GRB 190114C, in order to outline some of the intriguing features of long GRB associated to VHE energy afterglow observations. Under the guise of a proton synchrotron model, these features are as follows:
{\subsubsection{Particle index}}
{The parameters associated with $p_e \neq 
p_p$ yielded the most optimal results in these bursts. One potential cause of protons and electrons having distinct indices is the non-uniformity of the power-law turbulence spectrum of the magnetic field across a wide range of 
scales \citep{2009ApJ...699..953A}. This could be realized under several circumstances: a) a difference in scales corresponding to the gyration radii of protons attaining $10^{20}$eV, and electrons reaching GeV energies \citep{2015MNRAS.448..910C}, respectively, and b) a variation in wavelength distribution of shock generated magnetic perturbations and geometry of the magnetic field at the shock front  \citep{2006ApJ...650.1020N}.  
On the other hand, the acceleration processes setting these power-law indices in the presence of shock can have varied properties depending upon the orientation of the magnetic field relative to the shock \citep{2014ApJ...783...91C, 2014ApJ...794...47C, 2014ApJ...794..153G}.  
Apart from these, a two-component jet model could also feature different power law indices. In this model, a narrow, highly collimated jet could account for the TeV emission produced by protons, while a less collimated, wider jet could be responsible for the low-energy emission from electrons \citep{2003Natur.426..154B, 2004ApJ...605..300H, 2023MNRAS.522L..56S}.  }
\\
{\subsubsection{Embedded magnetic field}}
{Requiring that proton synchrotron emission explains the TeV observations results in the necessity of having a strong magnetic field, with $\epsilon_B$ close to equipartition. Similarly, the electron equipartition energy must be low, $\epsilon_e \lesssim 
10^{-3}$. Even though the inference $\epsilon_B \sim 10^{-1}$ is inconsistent with the analytical models and particle-in-cell (PIC) simulations of un-magnetized plasma \citep{2006ApJ...651L...9M, 2011ApJ...726...75S}, such a relatively high $\epsilon_B$ can be possibly explained as follows. 
Considering the connection between long-GRBs and supernovae could account for the highly magnetized environment as well as low  $\epsilon_e$ \citep[see \textit{e.g.}][]{1998ApJ...506L..27K, 2006A&A...447..121B, 
2006Natur.442.1008C, 2019A&A...622A.138K}. By performing global fitting of the emission in six supernova remnants (SNRs), \citet{2021ApJ...917...55R} found $\epsilon_B$ to be between $10^{-3}$ and $10^{-1}$, while $\epsilon_e$ achieves a smaller value in the range of 
$10^{-4} -  5 \cdot 10^{-2}$.  These equipartition parameters were determined 
owing to the advantageous conditions provided by SNRs. This is analogous to the findings in our model for 
this GRB and 190114C \citep{2022arXiv221002363I}. Also, the compression of the upstream turbulent magnetic field by 
the shock may amplifies its strength \citep{2006MNRAS.366..635L}. The type of turbulence spectrum may also dictate the strength of the magnetic field, such as Kolmogorov turbulence \citep{1987ApJ...322..643B} and Kraichnan turbulence \citep{1965PhFl....8.1385K}.  Alternatively, the reverse shock approximation could also be evoked for proton acceleration as it could harbor such a high magnetization 
\citep{2000ApJ...541..707W, 2018PhRvD..97h3010Z}. To explain the VHE emission of GRB 221009A with synchrotron emission from protons accelerated at the reverse shock, \citet{2022arXiv221105754Z} implemented a strong  magnetic field with 
$\epsilon_B \sim 0.5$.  }
\\
{\subsubsection{Injection fraction}}
The measured fluxes, together with the large kinetic energy, require the fraction of particle  accelerated into the power-law to be considerably low,
    $\xi_e \approx 10^{-2}$ and $\xi_p \approx 10^{-2}$. 
    These values are consistent with the numerical estimation 
    of $\xi_e \lesssim 10^{-1}$ considering the acceleration of charged particles in collisionless shocks \citep{2011ApJ...726...75S}. 
    A lower than the unity $\xi$ is gaining 
attention in the GRB afterglow theories and modeling, see e.g. 
\citet{2017ApJ...845..150R, 2018MNRAS.480.4060W, 
2020ApJ...904..166C, 2020ApJ...905..105A}. For instance, for the very nearby ($z\sim 0.07$), GRB 190829A, $\xi < 
1$ is obtained by a fit to the data \citep{2022ApJ...931L..19S}.
Also, \citet{2023arXiv230414331G} concluded their investigation of GRB 221009A by emphasizing the small value of $\xi_e \sim 10^{-2}$ required by their analysis. 
{Such a low injection fraction of electrons and protons indicate that there is a large population of thermal particle species present in the 
downstream of the shock \citep{2017ApJ...845..150R}. The thermal particles can indeed be anticipated to emit synchrotron radiation in radio frequencies during the early phase according to \citet{EW05}.
However, the contribution of these thermal electrons and protons in producing the GRB spectra is yet to be studied in detail \citep{2018MNRAS.480.4060W}.    } 
\\
{\subsubsection{Progenitor environment}}
{The progenitor of long-GRBs is a massive star which collapses to form a compact 
object \citep[see \textit{e.g.}][]{2015PhR...561....1K}. The low-metallicity Wolf-Rayet stars ($20 - 25 M_\odot$) with low mass-loss rates ($\sim 10^{-7} M_\odot \rm yr^{-1}$) are believed to be progenitors for the collapsar model for which the circumburst density is $n \gtrsim 10$~cm$^{-3}$ \citep{2002RvMP...74.1015W, 
2006ApJ...647.1269F}. Considering $\xi = 0.01$ in our model, the external medium density of these long duration bursts ranges between $10 - 10^2$~cm$^{-3}$ . When $\xi = 1$, for a constant ISM medium, the circumburst density estimated in theoretical models is around $n \leq 1$~cm$^{-3}$  \citep{2015MNRAS.454.1073B, 2018ApJ...866..162G, 2021ApJ...923..135D, 2023MNRAS.523..149G}, and can even reach $\sim 100$~cm$^{-3}$ \citep{2015ApJ...814....1L} in some specific cases.}
\\
{\subsubsection{Energetics of GRBs}}
For those two bursts, the required kinetic energy 
is larger by about an order of 
magnitude than their prompt-phase equivalent energy 
$E_{iso}$. This leads to a radiative efficiency 
$\eta = E_{iso}/(E_k+E_{iso})$ in the order of 10\%. For the proton synchrotron model, we estimated $\eta$ to be about $8\%$ for GRB 190114C and 9\% for GRB 221009A considering $p_e = 8/3$ and $p_p = 2.2$ for the latter. { Moreover, if the jet observed in GRB 221009A is considerably narrow with a half-opening angle of  $\theta \sim 0.8^\circ$ as reported by \citet{doi:10.1126/science.adg9328}, then the jet kinetic energy after correcting for beaming is $E_{k, jet} = (\theta^2/2)E_k \sim 3 \times 10^{51} $~erg for $p_e = 8/3$ and $p_p = 2.2$  while for $p_e = 8/3$ and $p_p = 2.3$ it is $E_{k, jet} \sim 5 \times 10^{51} $~erg. This is in agreement with the expectations of amount of energy stored in GRBs \citep{2001ApJ...562L..55F}. This analysis is based on the Konus-Wind estimation of $E_{iso} \sim 3 \times 10^{54}$~erg \citep{web:nasa:isotro}. However, \textit{Insight}-HXMT in conjunction with the GECAM-C measured the isotropic equivalent energy of GRB 221009A to be $1.5 \times 10^{55}$~erg \citep{2023arXiv230301203A}. The latter measurement is five times higher than the former. It is worthy to highlight that integrating $E_{iso} = 1.5 \times 10^{55}$~erg into our model leads to an increase in $\eta$ which is around $33 \%$ for $p_e = 8/3$ and $p_p = 2.2$ and $\sim 23\%$ considering $p_e = 8/3$ and $p_p = 2.3$,  while the other parameters remain unchanged. We emphasize that both estimates of $E_{iso}$ values yield reasonable prompt phase-energy conversion efficiencies. }
\\
{\subsubsection{Acceleration Efficiency}}
Finally, we found that these bursts do not require a very efficient proton acceleration, with an efficiency in the order of a few tens, $\alpha \approx 5 - 20$. Interestingly, the PIC simulations performed in  
\citet{2020ApJ...905..105A} in the context of the VHE afterglow emission of GRB 190114C resulted in even lower efficiency, $\alpha \sim 100$. {It is obtained by considering the early diffusive process in Fermi acceleration mechanism. The parameter $\alpha$ in the range $5 - 20$ acquired by setting the maximum proton energy may imply that high energy protons could be accelerated via MHD turbulence \citep{2018MNRAS.475.2713D, 2020ApJ...905..105A}.  }

\subsection{Comparison of SSC and proton-synchrotron components for GRB 221009A}

Many authors attempt to explain the VHE observations of GRB afterglow with a purely leptonic modeling based on the synchrotron self-Compton process \citep[for instance][]{2022arXiv221105754Z, 2022arXiv221015857G, 2022arXiv221010673R,2023arXiv230204388L, 2023arXiv230206225K,2023A&A...670L..12D, doi:10.1126/science.adg9328}. 
It is argued that explaining the high energy photon with this mechanism is difficult because the modelled SSC flux in the TeV band, which is strongly constrained by the radio, optical and X-ray fluxes,  is smaller than the corresponding observed flux after correcting for the EBL absorption \citep[see,][]{2022arXiv221015857G, 2022Galax..10...66M}. 
The obtained fits required lower magnetization than the one presented here, typically $\epsilon_B \approx 10^{-4} - 10^{-3}$ and higher $\epsilon_e \approx 10^{-2} - 10^{-1}$.

In the model we presented, this problem does not exist. Indeed the flux in the TeV band is somewhat independent from the flux at lower energies. However, this freedom comes at the expense of a large kinetic energy and a small fraction of electrons injected into the non-thermal power-law. This ultimately leads to a large external density for the interstellar medium and a small (5 to 10 \%) prompt radiative efficiency. We stress that these values are consistent with those found in numerical simulations, as well as afterglow modelling. 

\section{Conclusions}\label{sec:conclusion}

The explosions caused by the core collapse of massive stars are 
predicted to result in long-GRBs \citep[see \textit{e.g.}][]{2015PhR...561....1K}. Some of them, as 
identified recently, are accompanied by  VHE signals at energies $\gtrsim $ TeV during their 
afterglow phase. 
This offers an 
opportunity to investigate the source of VHE emission from these extremely powerful 
events.  We have explained 
the early afterglow of  GRB 221009A  within the framework of 
a hybrid emission model where the electron-synchrotron process is the source of the low energy component 
of the spectrum and the VHE component is explained by the 
proton-synchrotron mechanism with different particle indices. 
We constrain some parameters of this model by using observations in the optical, X-ray and TeV bands, and demonstrate that the observations can be reproduced by our model. Yet, our modeling requires that protons and electrons have different spectral indices. The key aspect of our model is that the kinetic energy of the bursts needs to be large and the fraction of particles (electrons and protons) accelerated into the power-law must be small.

We then compare the model parameters we obtained for GRB 221009A and for GRB 190114C \citep{2022arXiv221002363I} to underline their similarity. We find that explaining these two bursts with the hydrid model we presented requires a large kinetic energy, $E$ and density, $n$, which in turn limits the fractions of accelerated particles injected into the power-law by  shock 
acceleration to be small, with $\xi_e \sim 10^{-2}$ and $ \xi_p \gtrsim 10^{-2}$.
 Still, we emphasis the fact that the required energy in both cases, $\approx 10^{54}  - 10^{55}$~erg is not unacceptable. Especially for these extremely 
 bright GRBs, the efficiency in kinetic energy conversion to prompt emission is of the order of a few percent and up to 10\%. These values are not exceptional. Similar constraints are commonly inferred in GRBs, under various assumptions: for example in the context of purely leptonic models, see  \citet{2020ApJ...904..166C}. {The existence of a strong magnetic field, characterized by $\epsilon_B \gg \epsilon_e$ is crucial for a proton- synchrotron process to explain the TeV emission. We demonstrated that under this assumption, the high energy component of the SSC model experiences a significant suppression. However, the SSC process may take precedence in a scenario where $\epsilon_B \ll \epsilon_e$.}

We therefore conclude that the proton-synchrotron process offers a compelling 
alternative to other radiative models based on the SSC mechanisms to explain the VHE afterglow of the GRBs. Further detection of GRBs at VHE by the Cherenkov telescope array (CTA) \citep{knodlseder2020cherenkov} and the LHAASO experiment will allow to further constrain the free parameters and contrast it with other purely leptonic models.
\\
\\

We acknowledge support from the European Research Council via
ERC consolidating grant No. 773062 (acronym O.M.J.).

\vspace{5mm}

\appendix

\section{Inverse-Compton (IC) Scattering component } \label{app:IC}

First we point out that in our model we can adapt the
classical regime for the inverse Compton process,  \textit{i.e.} neglect
Klein-Nishina effects. Indeed 
\begin{align}
  \gamma_{m,e} h\nu_{m,e}\Gamma^{-1} &\sim 0.15~ E_{54}^{5/33} \epsilon_{p,-1}^{13/33} n_{0}^{5/33} \xi_p^{-1/11} ~\rm MeV.
\end{align}
Hence, the minimum frequency of the up-scattered electrons in the observer's frame of reference is given by $\nu_{m, IC}= 2 \gamma_{m,e}^2 \nu_{m,e}$. Using Equations \eqref{Eq:e_B} and \eqref{Eq:xi} at $t = 235$~s and $p_e = 2.67$, we find
\begin{align} \label{Eq:Ic_min_fqcy}
h\nu_{m, IC} &= 1.58 \times {{10}^{17}}\, {{0.025}^{\frac{4}{{p_e}-1}}} {{0.06}^{\frac{4 \left( {p_e}-2\right) }{{p_e}-1}}}\, {{6.44}^{\frac{2 {p_e}}{{p_e}-1}}}\, {({2.53 \cdot {{10}^{18}})}^{\frac{1-{p_p}}{{p_p}+1}}}\, {({8.94 \cdot {{10}^{-8}})}^{-\frac{2}{{p_p}+1}}}\, {{{{{f_p}}_p}}^{\frac{2 {p_p}-2}{{p_p}+1}}}\, \nonumber\\
&~~~ \times {{{E_{54}}}^{\frac{{p_p}+5}{4 {p_p}+4}}} {{{{{\epsilon}}_{p,-1}}}^{\frac{2 {p_p}-2}{{p_p}+1}}}\, {{{n_0}}^{\frac{3-{p_p}}{4 {p_p}+4}}}\, {{{{\xi}_p}}^{\frac{4-2 {p_p}}{{p_p}+1}}}\,,  \nonumber \\
&= 30~E_{54}^{73/132} \epsilon_{p,-1}^{26/33} n_0^{7/132}\xi_p^{-2/11} ~\text{~GeV}, 
\end{align}
Similarly, the characteristic cooling energy of the IC spectrum, $\nu_{c,IC}= 2 \gamma_{c,e}^2 \nu_{c,e}$, is given by 
\begin{align} \label{Eq:Ic_cool_fqcy}  
  h\nu_{c,IC} &= 1.55 \times {{10}^{8}}\, {({8.94 \cdot {{10}^{-8}})}^{-\frac{14}{{p_p}+1}}}\, {({2.53 \cdot {{10}^{18}})}^{-\frac{7 \left( {p_p}-1\right) }{{p_p}+1}}}\, {{{{{f_p}}_p}}^{\frac{14 {p_p}-14}{{p_p}+1}}}\, {{{E_{54}}}^{\frac{9 {p_p}+37}{4 {p_p}+4}}}\, {{{{{\epsilon}}_{p,-1}}}^{\frac{14 {p_p}-14}{{p_p}+1}}}\, \nonumber\\
  &~~~ \times {{n_0}}^{\frac{19-9{p_p}}{4 {p_p}+4}}\, {{{{\xi}_p}}^{\frac{28-14 {p_p}}{{p_p}+1}}}\, , \nonumber  \\
&= 6.8 \times 10^{-17} ~E_{54}^{577/132}\epsilon_{p,-1}^{182/33} n_{0}^{-17/132}\xi_p^{-14/11}\text{~eV}.
\end{align}
The maximum flux of the IC spectrum is given by $F_{\nu_{\rm peak, IC}} = \frac{1}{3}\sigma_T n_e r F_{\nu_{\rm peak, e}}$, and is expressed as
\begin{align}\label{Fmax_IC}
 F_{\nu_{\rm peak, IC}}  &= 3.51 \times {{10}^{-32}}\, {({2.53 \cdot {{10}^{18}})}^{\frac{2 {p_p}-2}{{p_p}+1}}}\, {({8.94 \cdot {{10}^{-8}})}^{\frac{4}{{p_p}+1}}}\, {{{{{f_p}}_p}}^{\frac{4-4 {p_p}}{{p_p}+1}}}\, {{{E_{54}}}^{\frac{-5 {p_p}-13}{4 {p_p}+4}}}\, {{{{{\epsilon}}_{p,-1}}}^{\frac{4-4 {p_p}}{{p_p}+1}}}\,  {{{n_0}}^{\frac{5 {p_p}-3}{4 {p_p}+4}}}\,  {{{{\xi}_p}}^{\frac{4 {p_p}-8}{{p_p}+1}}}, \nonumber \\
 &=3.2 \times 10^{-25}~E_{54}^{-245/132}\epsilon_{p,-1}^{-52/33} \xi_p^{4/11} n_0^{85/132} \text{~erg~cm$^{-2}$~s$^{-1}$~Hz$^{-1}$}.
\end{align}
Considering Equations \eqref{Eq:Ic_min_fqcy} and 
\eqref{Eq:Ic_cool_fqcy}, clearly $\nu_{m, IC} > \nu_{c, IC}$. 
The IC flux at 1 TeV for $p_e = 8/3$ and $t = 235$~s is then estimated to be
{
\begin{align}\label{Eq:flux_IC}
[\nu F_\nu]_{IC}|_{1 {\rm ~TeV}} &= 2.28 \times {{10}^{-5}}\, {{0.543}^{\frac{{p_e}}{2}}}\, {10}^{\frac{2-7p_e}{4}}\, {{{{{f_p}}_e}}^{2p_e-2}}\, {{{E_{54}}}^{\frac{3 {p_e}+2}{8}}} {{{{{\epsilon}}_{B,-1}}}^{\frac{{p_e}-6}{4}}} {{{{{\epsilon}}_{e,-2}}}^{2 {p_e}-2}}  {{{n_0}}^{\frac{2-{p_e}}{8}}}  {{{{\xi}_e}}^{4-2 {p_e}}}, \nonumber \\
&= 1.7 \times 10^{-10}
~ E_{54}^{5/4} \epsilon_{e,-2}^{10/3} \epsilon_{B,-1}^{-5/6} n_0^{-1/12} 
\xi_e^{-4/3}~~\text{erg~cm$^{-2}$~s$^{-1}$}.
\end{align}
}

\bibliography{sample631}{}

\begin{thebibliography}{}
\expandafter\ifx\csname natexlab\endcsname\relax\def\natexlab#1{#1}\fi
\providecommand{\url}[1]{\href{#1}{#1}}
\providecommand{\dodoi}[1]{doi:~\href{http://doi.org/#1}{\nolinkurl{#1}}}
\providecommand{\doeprint}[1]{\href{http://ascl.net/#1}{\nolinkurl{http://ascl.net/#1}}}
\providecommand{\doarXiv}[1]{\href{https://arxiv.org/abs/#1}{\nolinkurl{https://arxiv.org/abs/#1}}}

\bibitem[{{Abdalla} {et~al.}(2019){Abdalla}, {Adam}, {Aharonian}, {Ait
  Benkhali}, {Ang{\"u}ner}, {Arakawa}, {Arcaro}, {Armand}, {Ashkar}, {Backes},
  {Barbosa Martins}, {Barnard}, {Becherini}, {Berge}, {Bernl{\"o}hr},
  {Bissaldi}, {Blackwell}, {B{\"o}ttcher}, {Boisson}, {Bolmont}, {Bonnefoy},
  {Bregeon}, {Breuhaus}, {Brun}, {Brun}, {Bryan}, {B{\"u}chele}, {Bulik},
  {Bylund}, {Capasso}, {Caroff}, {Carosi}, {Casanova}, {Cerruti}, {Chand},
  {Chandra}, {Chen}, {Colafrancesco}, {Cury{\l}o}, {Davids}, {Deil}, {Devin},
  {deWilt}, {Dirson}, {Djannati-Ata{\"\i}}, {Dmytriiev}, {Donath},
  {Doroshenko}, {Dyks}, {Egberts}, {Emery}, {Ernenwein}, {Eschbach}, {Feijen},
  {Fegan}, {Fiasson}, {Fontaine}, {Funk}, {F{\"u}{\ss}ling}, {Gabici},
  {Gallant}, {Gat{\'e}}, {Giavitto}, {Giunti}, {Glawion}, {Glicenstein},
  {Gottschall}, {Grondin}, {Hahn}, {Haupt}, {Heinzelmann}, {Henri}, {Hermann},
  {Hinton}, {Hofmann}, {Hoischen}, {Holch}, {Holler}, {Horns}, {Huber},
  {Iwasaki}, {Jamrozy}, {Jankowsky}, {Jankowsky}, {Jardin-Blicq},
  {Jung-Richardt}, {Kastendieck}, {Katarzy{\'n}ski}, {Katsuragawa}, {Katz},
  {Khangulyan}, {Kh{\'e}lifi}, {King}, {Klepser}, {Klu{\'z}niak}, {Komin},
  {Kosack}, {Kostunin}, {Kreter}, {Lamanna}, {Lemi{\`e}re}, {Lemoine-Goumard},
  {Lenain}, {Leser}, {Levy}, {Lohse}, {Lypova}, {Mackey}, {Majumdar},
  {Malyshev}, {Marandon}, {Marcowith}, {Mares}, {Mariaud}, {Mart{\'\i}-Devesa},
  {Marx}, {Maurin}, {Meintjes}, {Mitchell}, {Moderski}, {Mohamed}, {Mohrmann},
  {Moore}, {Moulin}, {Muller}, {Murach}, {Nakashima}, {de Naurois},
  {Ndiyavala}, {Niederwanger}, {Niemiec}, {Oakes}, {O'Brien}, {Odaka}, {Ohm},
  {de Ona Wilhelmi}, {Ostrowski}, {Oya}, {Panter}, {Parsons}, {Perennes},
  {Petrucci}, {Peyaud}, {Piel}, {Pita}, {Poireau}, {Priyana Noel}, {Prokhorov},
  {Prokoph}, {P{\"u}hlhofer}, {Punch}, {Quirrenbach}, {Raab}, {Rauth},
  {Reimer}, {Reimer}, {Remy}, {Renaud}, {Rieger}, {Rinchiuso}, {Romoli},
  {Rowell}, {Rudak}, {Ruiz-Velasco}, {Sahakian}, {Sailer}, {Saito}, {Sanchez},
  {Santangelo}, {Sasaki}, {Schlickeiser}, {Sch{\"u}ssler}, {Schulz}, {Schutte},
  {Schwanke}, {Schwemmer}, {Seglar-Arroyo}, {Senniappan}, {Seyffert}, {Shafi},
  {Shiningayamwe}, {Simoni}, {Sinha}, {Sol}, {Specovius}, {Spir-Jacob},
  {Stawarz}, {Steenkamp}, {Stegmann}, {Steppa}, {Takahashi}, {Tavernier},
  {Taylor}, {Terrier}, {Tiziani}, {Tluczykont}, {Trichard}, {Tsirou}, {Tsuji},
  {Tuffs}, {Uchiyama}, {van der Walt}, {van Eldik}, {van Rensburg}, {van
  Soelen}, {Vasileiadis}, {Veh}, {Venter}, {Vincent}, {Vink}, {V{\"o}lk},
  {Vuillaume}, {Wadiasingh}, {Wagner}, {White}, {Wierzcholska}, {Yang},
  {Yoneda}, {Zacharias}, {Zanin}, {Zdziarski}, {Zech}, {Ziegler}, {Zorn},
  {{\.Z}ywucka}, {de Palma}, {Axelsson}, \& {Roberts}}]{2019Natur.575..464A}
{Abdalla}, H., {Adam}, R., {Aharonian}, F., {et~al.} 2019, \nat, 575, 464,
  \dodoi{10.1038/s41586-019-1743-9}

\bibitem[{Acciari {et~al.}(2019{\natexlab{a}})Acciari, Ansoldi, Antonelli,
  Engels, Baack, Babić, Banerjee, Barres~de Almeida, Barrio, González,
  Bednarek, Bellizzi, Bernardini, Berti, Besenrieder, Bhattacharyya,
  Bigongiari, Biland, Blanch, Bonnoli, Bošnjak, Busetto, Carosi, Ceribella,
  Chai, Chilingaryan, Cikota, Colak, Colin, Colombo, Contreras, Cortina,
  Covino, D’Elia, Da~Vela, Dazzi, De~Angelis, De~Lotto, Delfino, Delgado,
  Depaoli, Di~Pierro, Di~Venere, Espiñeira, Prester, Donini, Dorner, Doro,
  Elsaesser, Ramazani, Fattorini, Ferrara, Fidalgo, Foffano, Fonseca, Font,
  Fruck, Fukami, López, Garczarczyk, Gasparyan, Gaug, Giglietto, Giordano,
  Godinović, Green, Guberman, Hadasch, Hahn, Herrera, Hoang, Hrupec, Hütten,
  Inada, Inoue, Ishio, Iwamura, Jouvin, Kerszberg, Kubo, Kushida, Lamastra,
  Lelas, Leone, Lindfors, Lombardi, Longo, López, López-Coto, López-Oramas,
  Loporchio, de~Oliveira~Fraga, Maggio, Majumdar, Makariev, Mallamaci, Maneva,
  Manganaro, Mannheim, Maraschi, Mariotti, Martínez, Mazin, Mićanović,
  Miceli, Minev, Miranda, Mirzoyan, Molina, Moralejo, Morcuende, Moreno,
  Moretti, Munar-Adrover, Neustroev, Nigro, Nilsson, Ninci, Nishijima, Noda,
  Nogués, Nozaki, Paiano, Palatiello, Paneque, Paoletti, Paredes, Peñil,
  Peresano, Persic, Moroni, Prandini, Puljak, Rhode, Ribó, Rico, Righi,
  Rugliancich, Saha, Sahakyan, Saito, Sakurai, Satalecka, Schmidt, Schweizer,
  Sitarek, Šnidarić, Sobczynska, Somero, Stamerra, Strom, Strzys, Suda,
  Surić, Takahashi, Tavecchio, Temnikov, Terzić, Teshima, Torres-Albà,
  Tosti, Vagelli, van Scherpenberg, Vanzo, Acosta, Vigorito, Vitale, Vovk,
  Will, Zarić, Nava, Veres, Bhat, Briggs, Cleveland, Hamburg, Hui, Mailyan,
  Preece, Roberts, von Kienlin, Wilson-Hodge, Kocevski, Arimoto, Tak, Asano,
  Axelsson, Barbiellini, Bissaldi, Dirirsa, Gill, Granot, McEnery, Omodei,
  Razzaque, Piron, Racusin, Thompson, Campana, Bernardini, Kuin, Siegel, Cenko,
  O’Brien, Capalbi, Daì, De~Pasquale, Gropp, Klingler, Osborne, Perri,
  Starling, Tagliaferri, Tohuvavohu, Ursi, Tavani, Cardillo, Casentini, Piano,
  Evangelista, Verrecchia, Pittori, Lucarelli, Bulgarelli, Parmiggiani,
  Anderson, Anderson, Bernardi, Bolmer, Caballero-García, Carrasco,
  Castellón, Segura, Castro-Tirado, Cherukuri, Cockeram, D’Avanzo, Di~Dato,
  Diretse, Fender, Fernández-García, Fynbo, Fruchter, Greiner, Gromadzki,
  Heintz, Heywood, van~der Horst, Hu, Inserra, Izzo, Jaiswal, Jakobsson,
  Japelj, Kankare, Kann, Kouveliotou, Klose, Levan, Li, Lotti, Maguire,
  Malesani, Manulis, Marongiu, Martin, Melandri, Michałowski, Miller-Jones,
  Misra, Moin, Mooley, Nasri, Nicholl, Noschese, Novara, Pandey, Peretti, del
  Pulgar, Pérez-Torres, Perley, Piro, Ragosta, Resmi, Ricci, Rossi,
  Sánchez-Ramírez, Selsing, Schulze, Smartt, Smith, Sokolov, Stevens, Tanvir,
  Thöne, Tiengo, Tremou, Troja, de~Ugarte~Postigo, \& {MAGIC
  Collaboration}}]{acciari_observation_2019}
Acciari, V.~A., Ansoldi, S., Antonelli, L.~A., {et~al.} 2019{\natexlab{a}},
  Nature, 575, 459, \dodoi{10.1038/s41586-019-1754-6}

\bibitem[{Acciari {et~al.}(2019{\natexlab{b}})}]{MAGIC:2019irs}
Acciari, V.~A., {et~al.} 2019{\natexlab{b}}, Nature, 575, 459,
  \dodoi{10.1038/s41586-019-1754-6}

\bibitem[{Ackermann {et~al.}(2012)Ackermann, Ajello, Allafort, Schady, Baldini,
  Ballet, Barbiellini, Bastieri, Bellazzini, Blandford, Bloom, Borgland,
  Bottacini, Bouvier, Bregeon, Brigida, Bruel, Buehler, Buson, Caliandro,
  Cameron, Caraveo, Cavazzuti, Cecchi, Charles, Chaves, Chekhtman, Cheung,
  Chiang, Chiaro, Ciprini, Claus, Cohen-Tanugi, Conrad, Cutini, D’Ammando,
  de~Palma, Dermer, Digel, do~Couto~e Silva, Domínguez, Drell, Drlica-Wagner,
  Favuzzi, Fegan, Focke, Franckowiak, Fukazawa, Funk, Fusco, Gargano,
  Gasparrini, Gehrels, Germani, Giglietto, Giordano, Giroletti, Glanzman,
  Godfrey, Grenier, Grove, Guiriec, Gustafsson, Hadasch, Hayashida, Hays,
  Jackson, Jogler, Kataoka, Knödlseder, Kuss, Lande, Larsson, Latronico,
  Longo, Loparco, Lovellette, Lubrano, Mazziotta, McEnery, Mehault, Michelson,
  Mizuno, Monte, Monzani, Morselli, Moskalenko, Murgia, Tramacere, Nuss,
  Greiner, Ohno, Ohsugi, Omodei, Orienti, Orlando, Ormes, Paneque, Perkins,
  Pesce-Rollins, Piron, Pivato, Porter, Rainò, Rando, Razzano, Razzaque,
  Reimer, Reimer, Reyes, Ritz, Rau, Romoli, Roth, Sánchez-Conde, Sanchez,
  Scargle, Sgrò, Siskind, Spandre, Spinelli, Łukasz Stawarz, Suson,
  Takahashi, Tanaka, Thayer, Thompson, Tibaldo, Tinivella, Torres, Tosti,
  Troja, Usher, Vandenbroucke, Vasileiou, Vianello, Vitale, Waite, Winer, Wood,
  \& Wood}]{doi:10.1126/science.1227160}
Ackermann, M., Ajello, M., Allafort, A., {et~al.} 2012, Science, 338, 1190,
  \dodoi{10.1126/science.1227160}

\bibitem[{{Aharonian} {et~al.}(1997){Aharonian}, {Hofmann}, {Konopelko}, \&
  {V{\"o}lk}}]{1997APh.....6..369A}
{Aharonian}, F.~A., {Hofmann}, W., {Konopelko}, A.~K., \& {V{\"o}lk}, H.~J.
  1997, Astroparticle Physics, 6, 369, \dodoi{10.1016/S0927-6505(96)00070-9}

\bibitem[{{An} {et~al.}(2023){An}, {Antier}, {Bi}, {Bu}, {Cai}, {Cao},
  {Camisasca}, {Chang}, {Chen}, {Chen}, {Chen}, {Chen}, {Chen}, {Chen}, {Chen},
  {Coughlin}, {Cui}, {Dai}, {Hussenot-Desenonges}, {Du}, {Du}, {Du}, {Fan},
  {Frontera}, {Gao}, {Gao}, {Ge}, {Gong}, {Gu}, {Guan}, {Guo}, {Guo},
  {Guidorzi}, {Han}, {He}, {He}, {Hou}, {Huang}, {Huo}, {Ji}, {Jia}, {Jiang},
  {Kann}, {Klotz}, {Kong}, {Lan}, {Li}, {Li}, {Li}, {Li}, {Li}, {Li}, {Li},
  {Li}, {Li}, {Li}, {Li}, {Li}, {Li}, {Liang}, {Liang}, {Liao}, {Lin}, {Liu},
  {Liu}, {Liu}, {Liu}, {Liu}, {Liu}, {Liu}, {Lu}, {Lu}, {Lu}, {Luo}, {Luo},
  {Ma}, {Ma}, {Ma}, {Ma}, {Maccary}, {Mao}, {Meng}, {Nie}, {Orlandini}, {Ou},
  {Peng}, {Peng}, {Qiao}, {Qu}, {Ren}, {Shi}, {Shi}, {Song}, {Song}, {Su},
  {Sun}, {Sun}, {Sun}, {Tan}, {Tan}, {Tao}, {Tuo}, {Turpin}, {Wang}, {Wang},
  {Wang}, {Wang}, {Wang}, {Wang}, {Wang}, {Wang}, {Wang}, {Wang}, {Wang},
  {Wang}, {Wang}, {Wang}, {Wen}, {Wu}, {Wu}, {Wu}, {Xiao}, {Xiao}, {Xiao},
  {Xie}, {Xiong}, {Xiong}, {Xu}, {Xu}, {Xu}, {Xu}, {Xu}, {Xu}, {Xue}, {Yang},
  {Yang}, {Yang}, {Ye}, {Yi}, {Yi}, {Yin}, {You}, {Yu}, {Yu}, {Yu}, {Zeng},
  {Zhang}, {Zhang}, {Zhang}, {Zhang}, {Zhang}, {Zhang}, {Zhang}, {Zhang},
  {Zhang}, {Zhang}, {Zhang}, {Zhang}, {Zhang}, {Zhang}, {Zhang}, {Zhang},
  {Zhang}, {Zhang}, {Zhang}, {Zhao}, {Zhao}, {Zhao}, {Zhao}, {Zhao}, {Zhao},
  {Zhao}, {Zhao}, {Zheng}, {Zheng}, {Zhou}, {Zhou}, \&
  {Zhu}}]{2023arXiv230301203A}
{An}, Z.-H., {Antier}, S., {Bi}, X.-Z., {et~al.} 2023, arXiv e-prints,
  arXiv:2303.01203, \dodoi{10.48550/arXiv.2303.01203}

\bibitem[{{Asano} {et~al.}(2009){Asano}, {Inoue}, \&
  {M{\'e}sz{\'a}ros}}]{2009ApJ...699..953A}
{Asano}, K., {Inoue}, S., \& {M{\'e}sz{\'a}ros}, P. 2009, \apj, 699, 953,
  \dodoi{10.1088/0004-637X/699/2/953}

\bibitem[{{Asano} {et~al.}(2020){Asano}, {Murase}, \&
  {Toma}}]{2020ApJ...905..105A}
{Asano}, K., {Murase}, K., \& {Toma}, K. 2020, \apj, 905, 105,
  \dodoi{10.3847/1538-4357/abc82c}

\bibitem[{{Beniamini} {et~al.}(2015){Beniamini}, {Nava}, {Duran}, \&
  {Piran}}]{2015MNRAS.454.1073B}
{Beniamini}, P., {Nava}, L., {Duran}, R.~B., \& {Piran}, T. 2015, \mnras, 454,
  1073, \dodoi{10.1093/mnras/stv2033}

\bibitem[{{Beniamini} {et~al.}(2016){Beniamini}, {Nava}, \&
  {Piran}}]{2016MNRAS.461...51B}
{Beniamini}, P., {Nava}, L., \& {Piran}, T. 2016, \mnras, 461, 51,
  \dodoi{10.1093/mnras/stw1331}

\bibitem[{{Berger} {et~al.}(2003){Berger}, {Kulkarni}, {Pooley}, {Frail},
  {McIntyre}, {Wark}, {Sari}, {Soderberg}, {Fox}, {Yost}, \&
  {Price}}]{2003Natur.426..154B}
{Berger}, E., {Kulkarni}, S.~R., {Pooley}, G., {et~al.} 2003, \nat, 426, 154,
  \dodoi{10.1038/nature01998}

\bibitem[{{Biermann} \& {Strittmatter}(1987)}]{1987ApJ...322..643B}
{Biermann}, P.~L., \& {Strittmatter}, P.~A. 1987, \apj, 322, 643,
  \dodoi{10.1086/165759}

\bibitem[{{Blanch} {et~al.}(2020){Blanch}, {Longo}, {Berti}, {Fukami}, {Suda},
  {Loporchio}, {Micanovic}, {Green}, {Pinter}, {Takahashi}, \& {MAGIC
  Collaboration}}]{2020GCN.29075....1B}
{Blanch}, O., {Longo}, F., {Berti}, A., {et~al.} 2020, GRB Coordinates Network,
  29075, 1

\bibitem[{{Blandford} \& {McKee}(1976)}]{Blandford:1976}
{Blandford}, R.~D., \& {McKee}, C.~F. 1976, Physics of Fluids, 19, 1130,
  \dodoi{10.1063/1.861619}

\bibitem[{{Bosnjak} {et~al.}(2006){Bosnjak}, {Celotti}, {Ghirlanda}, {Della
  Valle}, \& {Pian}}]{2006A&A...447..121B}
{Bosnjak}, Z., {Celotti}, A., {Ghirlanda}, G., {Della Valle}, M., \& {Pian}, E.
  2006, \aap, 447, 121, \dodoi{10.1051/0004-6361:20052803}

\bibitem[{{B{\"o}ttcher} \& {Dermer}(1998)}]{1998ApJ...499L.131B}
{B{\"o}ttcher}, M., \& {Dermer}, C.~D. 1998, \apjl, 499, L131,
  \dodoi{10.1086/311366}

\bibitem[{Cai {et~al.}(2021)}]{Cai:2021lmr}
Cai, C., {et~al.} 2021, Mon. Not. Roy. Astron. Soc., 508, 3910,
  \dodoi{10.1093/mnras/stab2760}

\bibitem[{{Campana} {et~al.}(2006){Campana}, {Mangano}, {Blustin}, {Brown},
  {Burrows}, {Chincarini}, {Cummings}, {Cusumano}, {Della Valle}, {Malesani},
  {M{\'e}sz{\'a}ros}, {Nousek}, {Page}, {Sakamoto}, {Waxman}, {Zhang}, {Dai},
  {Gehrels}, {Immler}, {Marshall}, {Mason}, {Moretti}, {O'Brien}, {Osborne},
  {Page}, {Romano}, {Roming}, {Tagliaferri}, {Cominsky}, {Giommi}, {Godet},
  {Kennea}, {Krimm}, {Angelini}, {Barthelmy}, {Boyd}, {Palmer}, {Wells}, \&
  {White}}]{2006Natur.442.1008C}
{Campana}, S., {Mangano}, V., {Blustin}, A.~J., {et~al.} 2006, \nat, 442, 1008,
  \dodoi{10.1038/nature04892}

\bibitem[{{Cao} {et~al.}(2019){Cao}, {della Volpe}, {Liu}, {Editors}, {:},
  {Bi}, {Chen}, {D'Ettorre Piazzoli}, {Feng}, {Jia}, {Li}, {Ma}, {Wang},
  {Zhang}, {Referees}, {:}, {Qie}, {Hu}, {Referees}, {:}, {S{\'a}iz}, {Yang},
  {Contributors}, {:}, {Addazi}, {Belotsky}, {Beylin}, {Bi}, {Che}, {Chen},
  {Cheng}, {Chiavassa}, {Cirelli}, {Di Sciascio}, {Esmaili}, {Fang},
  {Fornengo}, {Gou}, {Guo}, {Gan}, {Gong}, {Gu}, {He}, {He}, {Hou}, {Huang},
  {Huang}, {Kachekriess}, {Khlopov}, {Korchagin}, {Korochkin}, {Kuksa},
  {Ksenofontov}, {Liu}, {Liu}, {Liu}, {Marciano}, {Martineau-Huynh},
  {Martraire}, {Ma}, {Neronov}, {Panci}, {Pasechnick}, {Ruffolo}, {Sakharov},
  {Sala}, {Semikoz}, {Shchegolev}, {Serpico}, {Sheng}, {Stenkin}, {Tam},
  {Vernetto}, {Vallania}, {Volchanskiy}, {Wang}, {Wang}, {Wang}, {Wu}, {Wu},
  {Wu}, {Xiao}, {Yang}, {Yan}, {Yao}, {Yin}, {Yuan}, {Zhang}, {Zeng}, {Zhang},
  {Zhang}, {Zhou}, {Zhu}, \& {Zuo}}]{2019arXiv190502773C}
{Cao}, Z., {della Volpe}, D., {Liu}, S., {et~al.} 2019, arXiv e-prints,
  arXiv:1905.02773.
\newblock \doarXiv{1905.02773}

\bibitem[{{Caprioli} \& {Spitkovsky}(2014{\natexlab{a}})}]{2014ApJ...783...91C}
{Caprioli}, D., \& {Spitkovsky}, A. 2014{\natexlab{a}}, \apj, 783, 91,
  \dodoi{10.1088/0004-637X/783/2/91}

\bibitem[{{Caprioli} \& {Spitkovsky}(2014{\natexlab{b}})}]{2014ApJ...794...47C}
---. 2014{\natexlab{b}}, \apj, 794, 47, \dodoi{10.1088/0004-637X/794/1/47}

\bibitem[{{Castro-Tirado} {et~al.}(2022){Castro-Tirado}, {Sanchez-Ramirez},
  {Hu}, {Caballero-Garcia}, {Castro Tirado}, {Fernandez-Garcia},
  {Perez-Garcia}, {Lombardi}, {Pandey}, {Yang}, \&
  {Zhang}}]{2022GCN.32686....1C}
{Castro-Tirado}, A.~J., {Sanchez-Ramirez}, R., {Hu}, Y.~D., {et~al.} 2022, GRB
  Coordinates Network, 32686, 1

\bibitem[{{Cerruti} {et~al.}(2015){Cerruti}, {Zech}, {Boisson}, \&
  {Inoue}}]{2015MNRAS.448..910C}
{Cerruti}, M., {Zech}, A., {Boisson}, C., \& {Inoue}, S. 2015, \mnras, 448,
  910, \dodoi{10.1093/mnras/stu2691}

\bibitem[{{Cunningham} {et~al.}(2020){Cunningham}, {Cenko}, {Ryan}, {Vogel},
  {Corsi}, {Cucchiara}, {Fruchter}, {Horesh}, {Kangas}, {Kocevski}, {Perley},
  \& {Racusin}}]{2020ApJ...904..166C}
{Cunningham}, V., {Cenko}, S.~B., {Ryan}, G., {et~al.} 2020, \apj, 904, 166,
  \dodoi{10.3847/1538-4357/abc2cd}

\bibitem[{{Das} \& {Razzaque}(2023)}]{2023A&A...670L..12D}
{Das}, S., \& {Razzaque}, S. 2023, \aap, 670, L12,
  \dodoi{10.1051/0004-6361/202245377}

\bibitem[{de~Ugarte Postigo~A. {et~al.}(2022)de~Ugarte Postigo~A., Izzo, \&
  Pugliese}]{web:nasa:redshift}
de~Ugarte Postigo~A., Izzo, L., \& Pugliese, G. e.~a. 2022, GRB 221009A:
  Redshift from X-shooter/VLT.
\newblock \url{https://gcn.gsfc.nasa.gov/gcn3/32648.gcn3}

\bibitem[{{Demidem} {et~al.}(2018){Demidem}, {Lemoine}, \&
  {Casse}}]{2018MNRAS.475.2713D}
{Demidem}, C., {Lemoine}, M., \& {Casse}, F. 2018, \mnras, 475, 2713,
  \dodoi{10.1093/mnras/stx3367}

\bibitem[{{Derishev} \& {Piran}(2021{\natexlab{a}})}]{Derishev:2021ivd}
{Derishev}, E., \& {Piran}, T. 2021{\natexlab{a}}, arXiv e-prints,
  arXiv:2106.12035.
\newblock \doarXiv{2106.12035}

\bibitem[{{Derishev} \& {Piran}(2021{\natexlab{b}})}]{2021ApJ...923..135D}
---. 2021{\natexlab{b}}, \apj, 923, 135, \dodoi{10.3847/1538-4357/ac2dec}

\bibitem[{Dermer {et~al.}(2000)Dermer, Chiang, \& Mitman}]{Dermer_2000}
Dermer, C.~D., Chiang, J., \& Mitman, K.~E. 2000, The Astrophysical Journal,
  537, 785, \dodoi{10.1086/309061}

\bibitem[{{Dichiara} {et~al.}(2022){Dichiara}, {Gropp}, {Kennea}, {Kuin},
  {Lien}, {Marshall}, {Tohuvavohu}, \& {Williams}}]{2022ATel15650....1D}
{Dichiara}, S., {Gropp}, J.~D., {Kennea}, J.~A., {et~al.} 2022, The
  Astronomer's Telegram, 15650, 1

\bibitem[{{Eichler} \& {Waxman}(2005)}]{EW05}
{Eichler}, D., \& {Waxman}, E. 2005, \apj, 627, 861, \dodoi{10.1086/430596}

\bibitem[{{Evans} {et~al.}(2007){Evans}, {Beardmore}, {Page}, {Tyler},
  {Osborne}, {Goad}, {O'Brien}, {Vetere}, {Racusin}, {Morris}, {Burrows},
  {Capalbi}, {Perri}, {Gehrels}, \& {Romano}}]{2007A&A...469..379E}
{Evans}, P.~A., {Beardmore}, A.~P., {Page}, K.~L., {et~al.} 2007, \aap, 469,
  379, \dodoi{10.1051/0004-6361:20077530}

\bibitem[{{Evans} {et~al.}(2009){Evans}, {Beardmore}, {Page}, {Osborne},
  {O'Brien}, {Willingale}, {Starling}, {Burrows}, {Godet}, {Vetere}, {Racusin},
  {Goad}, {Wiersema}, {Angelini}, {Capalbi}, {Chincarini}, {Gehrels}, {Kennea},
  {Margutti}, {Morris}, {Mountford}, {Pagani}, {Perri}, {Romano}, \&
  {Tanvir}}]{2009MNRAS.397.1177E}
---. 2009, \mnras, 397, 1177, \dodoi{10.1111/j.1365-2966.2009.14913.x}

\bibitem[{{Fraija} {et~al.}(2022){Fraija}, {Dainotti}, {Ugale}, {Jyoti}, \&
  {Warren}}]{2022ApJ...934..188F}
{Fraija}, N., {Dainotti}, M.~G., {Ugale}, S., {Jyoti}, D., \& {Warren}, D.~C.
  2022, \apj, 934, 188, \dodoi{10.3847/1538-4357/ac7a9c}

\bibitem[{Fraija {et~al.}(2019)Fraija, Duran, Dichiara, \&
  Beniamini}]{fraija2019synchrotron}
Fraija, N., Duran, R.~B., Dichiara, S., \& Beniamini, P. 2019, The
  Astrophysical Journal, 883, 162

\bibitem[{{Frail} {et~al.}(2001){Frail}, {Kulkarni}, {Sari}, {Djorgovski},
  {Bloom}, {Galama}, {Reichart}, {Berger}, {Harrison}, {Price}, {Yost},
  {Diercks}, {Goodrich}, \& {Chaffee}}]{2001ApJ...562L..55F}
{Frail}, D.~A., {Kulkarni}, S.~R., {Sari}, R., {et~al.} 2001, \apjl, 562, L55,
  \dodoi{10.1086/338119}

\bibitem[{{Franceschini}(2021)}]{2021Univ....7..146F}
{Franceschini}, A. 2021, Universe, 7, 146, \dodoi{10.3390/universe7050146}

\bibitem[{Frederiks {et~al.}(2022)Frederiks, A.Lysenko, Ridnaia, Svinkin,
  Tsvetkova, Ulanov, \& Cline}]{web:nasa:isotro}
Frederiks, D., A.Lysenko, Ridnaia, A., {et~al.} 2022, Konus-Wind detection of
  GRB 221009A.
\newblock \url{https://gcn.gsfc.nasa.gov/gcn3/32668.gcn3}

\bibitem[{{Frederiks} {et~al.}(2023){Frederiks}, {Svinkin}, {Lysenko},
  {Molkov}, {Tsvetkova}, {Ulanov}, {Ridnaia}, {Lutovinov}, {Lapshov},
  {Tkachenko}, \& {Levin}}]{2023arXiv230213383F}
{Frederiks}, D., {Svinkin}, D., {Lysenko}, A.~L., {et~al.} 2023, arXiv
  e-prints, arXiv:2302.13383, \dodoi{10.48550/arXiv.2302.13383}

\bibitem[{{Fryer} {et~al.}(2006){Fryer}, {Rockefeller}, \&
  {Young}}]{2006ApJ...647.1269F}
{Fryer}, C.~L., {Rockefeller}, G., \& {Young}, P.~A. 2006, \apj, 647, 1269,
  \dodoi{10.1086/505590}

\bibitem[{{Fulton} {et~al.}(2023){Fulton}, {Smartt}, {Rhodes}, {Huber},
  {Villar}, {Moore}, {Srivastav}, {Schultz}, {Chambers}, {Izzo}, {Hjorth},
  {Chen}, {Nicholl}, {Foley}, {Rest}, {Smith}, {Young}, {Sim}, {Bright},
  {Zenati}, {de Boer}, {Bulger}, {Fairlamb}, {Gao}, {Lin}, {Lowe}, {Magnier},
  {Smith}, {Wainscoat}, {Coulter}, {Jones}, {Kilpatrick}, {McGill},
  {Ramirez-Ruiz}, {Lee}, {Narayan}, {Ramakrishnan}, {Ridden-Harper}, {Singh},
  {Wang}, {Kong}, {Ngeow}, {Pan}, {Yang}, {Davis}, {Piro}, {Rojas-Bravo},
  {Sommer}, \& {Yadavalli}}]{2023ApJ...946L..22F}
{Fulton}, M.~D., {Smartt}, S.~J., {Rhodes}, L., {et~al.} 2023, \apjl, 946, L22,
  \dodoi{10.3847/2041-8213/acc101}

\bibitem[{Ghisellini \& Celotti(1998)}]{ghisellini1998quasi}
Ghisellini, G., \& Celotti, A. 1998, The Astrophysical Journal, 511, L93

\bibitem[{{Gill} \& {Granot}(2023)}]{2023arXiv230414331G}
{Gill}, R., \& {Granot}, J. 2023, arXiv e-prints, arXiv:2304.14331,
  \dodoi{10.48550/arXiv.2304.14331}

\bibitem[{{Gompertz} {et~al.}(2018){Gompertz}, {Fruchter}, \&
  {Pe'er}}]{2018ApJ...866..162G}
{Gompertz}, B.~P., {Fruchter}, A.~S., \& {Pe'er}, A. 2018, \apj, 866, 162,
  \dodoi{10.3847/1538-4357/aadba8}

\bibitem[{{Gonz{\'a}lez} {et~al.}(2022){Gonz{\'a}lez}, {Avila Rojas}, {Pratts},
  {Hern{\'a}ndez-Cadena}, {Fraija}, {Alfaro}, {P{\'e}rez Araujo}, \&
  {Montes}}]{2022arXiv221015857G}
{Gonz{\'a}lez}, M.~M., {Avila Rojas}, D., {Pratts}, A., {et~al.} 2022, arXiv
  e-prints, arXiv:2210.15857, \dodoi{10.48550/arXiv.2210.15857}

\bibitem[{{Guarini} {et~al.}(2023){Guarini}, {Tamborra}, {B{\'e}gu{\'e}}, \&
  {Rudolph}}]{2023MNRAS.523..149G}
{Guarini}, E., {Tamborra}, I., {B{\'e}gu{\'e}}, D., \& {Rudolph}, A. 2023,
  \mnras, 523, 149, \dodoi{10.1093/mnras/stad1421}

\bibitem[{{Guo} {et~al.}(2014){Guo}, {Sironi}, \&
  {Narayan}}]{2014ApJ...794..153G}
{Guo}, X., {Sironi}, L., \& {Narayan}, R. 2014, \apj, 794, 153,
  \dodoi{10.1088/0004-637X/794/2/153}

\bibitem[{{H.~E.~S.~S. Collaboration} {et~al.}(2021){H.~E.~S.~S.
  Collaboration}, {Abdalla}, {Aharonian}, {Ait Benkhali}, {Ang{\"u}ner},
  {Arcaro}, {Armand}, {Armstrong}, {Ashkar}, {Backes}, {Baghmanyan}, {Barbosa
  Martins}, {Barnacka}, {Barnard}, {Becherini}, {Berge}, {Bernl{\"o}hr}, {Bi},
  {Bissaldi}, {B{\"o}ttcher}, {Boisson}, {Bolmont}, {de Bony de Lavergne},
  {Breuhaus}, {Brun}, {Brun}, {Bryan}, {B{\"u}chele}, {Bulik}, {Bylund},
  {Caroff}, {Carosi}, {Casanova}, {Chand}, {Chandra}, {Chen}, {Cotter},
  {Cury{\l}o}, {Damascene Mbarubucyeye}, {Davids}, {Davies}, {Deil}, {Devin},
  {Dirson}, {Djannati-Ata{\"\i}}, {Dmytriiev}, {Donath}, {Doroshenko},
  {Dreyer}, {Duffy}, {Dyks}, {Egberts}, {Eichhorn}, {Einecke}, {Emery},
  {Ernenwein}, {Feijen}, {Fegan}, {Fiasson}, {Fichet de Clairfontaine},
  {Fontaine}, {Funk}, {F{\"u}{\ss}ling}, {Gabici}, {Gallant}, {Giavitto},
  {Giunti}, {Glawion}, {Glicenstein}, {Grondin}, {Hahn}, {Haupt}, {Hermann},
  {Hinton}, {Hofmann}, {Hoischen}, {Holch}, {Holler}, {H{\"o}rbe}, {Horns},
  {Huber}, {Jamrozy}, {Jankowsky}, {Jankowsky}, {Jardin-Blicq}, {Joshi},
  {Jung-Richardt}, {Kasai}, {Kastendieck}, {Katarzy{\'n}ski}, {Katz},
  {Khangulyan}, {Kh{\'e}lifi}, {Klepser}, {Klu{\'z}niak}, {Komin}, {Konno},
  {Kosack}, {Kostunin}, {Kreter}, {Lamanna}, {Lemi{\`e}re}, {Lemoine-Goumard},
  {Lenain}, {Leuschner}, {Levy}, {Lohse}, {Lypova}, {Mackey}, {Majumdar},
  {Malyshev}, {Malyshev}, {Marandon}, {Marchegiani}, {Marcowith}, {Mares},
  {Mart{\'\i}-Devesa}, {Marx}, {Maurin}, {Meintjes}, {Meyer}, {Mitchell},
  {Moderski}, {Mohrmann}, {Montanari}, {Moore}, {Morris}, {Moulin}, {Muller},
  {Murach}, {Nakashima}, {Nayerhoda}, {de Naurois}, {Ndiyavala}, {Niemiec},
  {Oakes}, {O'Brien}, {Odaka}, {Ohm}, {Olivera-Nieto}, {de Ona Wilhelmi},
  {Ostrowski}, {Panny}, {Panter}, {Parsons}, {Peron}, {Peyaud}, {Piel}, {Pita},
  {Poireau}, {Priyana Noel}, {Prokhorov}, {Prokoph}, {P{\"u}hlhofer}, {Punch},
  {Quirrenbach}, {Raab}, {Rauth}, {Reichherzer}, {Reimer}, {Reimer}, {Remy},
  {Renaud}, {Rieger}, {Rinchiuso}, {Romoli}, {Rowell}, {Rudak}, {Ruiz-Velasco},
  {Sahakian}, {Sailer}, {Salzmann}, {Sanchez}, {Santangelo}, {Sasaki},
  {Scalici}, {Sch{\"a}fer}, {Sch{\"u}ssler}, {Schutte}, {Schwanke},
  {Seglar-Arroyo}, {Senniappan}, {Seyffert}, {Shafi}, {Shapopi},
  {Shiningayamwe}, {Simoni}, {Sinha}, {Sol}, {Specovius}, {Spencer},
  {Spir-Jacob}, {Stawarz}, {Sun}, {Steenkamp}, {Stegmann}, {Steinmassl},
  {Steppa}, {Takahashi}, {Tam}, {Tavernier}, {Taylor}, {Terrier}, {Thiersen},
  {Tiziani}, {Tluczykont}, {Tomankova}, {Tsirou}, {Tuffs}, {Uchiyama}, {van der
  Walt}, {van Eldik}, {van Rensburg}, {van Soelen}, {Vasileiadis}, {Veh},
  {Venter}, {Vincent}, {Vink}, {V{\"o}lk}, {Wadiasingh}, {Wagner}, {Watson},
  {Werner}, {White}, {Wierzcholska}, {Wong}, {Yusafzai}, {Zacharias}, {Zanin},
  {Zargaryan}, {Zdziarski}, {Zech}, {Zhu}, {Zorn}, {Zouari}, {{\.Z}ywucka},
  {Evans}, \& {Page}}]{2021Sci...372.1081H}
{H.~E.~S.~S. Collaboration}, {Abdalla}, H., {Aharonian}, F., {et~al.} 2021,
  Science, 372, 1081, \dodoi{10.1126/science.abe8560}

\bibitem[{Hu {et~al.}(2022)Hu, Casanova, \& Fernandez-Garcia}]{web:nasa:Rband}
Hu, Y.-D., Casanova, V., \& Fernandez-Garcia, E., e.~a. 2022, GRB 221009A
  BOOTES-2/TELMA and OSN optical detections.
\newblock \url{https://gcn.gsfc.nasa.gov/gcn3/32644.gcn3}

\bibitem[{{Huang} {et~al.}(2004){Huang}, {Wu}, {Dai}, {Ma}, \&
  {Lu}}]{2004ApJ...605..300H}
{Huang}, Y.~F., {Wu}, X.~F., {Dai}, Z.~G., {Ma}, H.~T., \& {Lu}, T. 2004, \apj,
  605, 300, \dodoi{10.1086/382202}

\bibitem[{Huang.~Y.(2022)}]{web:nasa:LHAASO}
Huang.~Y., Hu.~S., . S. C.~e. 2022, LHAASO observed GRB 221009A with more than
  5000 VHE photons up to around 18 TeV.
\newblock \url{https://gcn.gsfc.nasa.gov/gcn3/32677.gcn3}

\bibitem[{{Isravel} {et~al.}(2022){Isravel}, {Pe'er}, \&
  {Begue}}]{2022arXiv221002363I}
{Isravel}, H., {Pe'er}, A., \& {Begue}, D. 2022, arXiv e-prints,
  arXiv:2210.02363, \dodoi{10.48550/arXiv.2210.02363}

\bibitem[{{Kann} {et~al.}(2023){Kann}, {Agayeva}, {Aivazyan}, {Alishov},
  {Andrade}, {Antier}, {Baransky}, {Bendjoya}, {Benkhaldoun}, {Beradze},
  {Berezin}, {Bo{\"e}r}, {Broens}, {Brunier}, {Bulla}, {Burkhonov}, {Burns},
  {Chen}, {Chen}, {Conti}, {Coughlin}, {Cui}, {Daigne}, {Delaveau},
  {Devillepoix}, {Dietrich}, {Dornic}, {Dubois}, {Ducoin}, {Durand}, {Duverne},
  {Eggenstein}, {Ehgamberdiev}, {Fouad}, {Freeberg}, {Froebrich}, {Ge},
  {Gervasoni}, {Godunova}, {Gokuldass}, {Gurbanov}, {Han}, {Hasanov}, {Hello},
  {Hussenot-Desenonges}, {Inasaridze}, {Iskandar}, {Ismailov}, {Janati}, {Jegou
  du Laz}, {Jia}, {Karpov}, {Kaeouach}, {Kiendrebeogo}, {Klotz}, {Kneip},
  {Kochiashvili}, {Kunert}, {Lekic}, {Leonini}, {Li}, {Li}, {Li}, {Liao},
  {Logie}, {Lu}, {Mao}, {Marchais}, {M{\'e}nard}, {Morris}, {Natsvlishvili},
  {Nedora}, {Noonan}, {Noysena}, {Orange}, {Pang}, {Peng}, {Pellouin},
  {Peloton}, {Pradier}, {Pyshna}, {Rajabo}, {Rau}, {Rinner}, {Rivet},
  {Romanov}, {Rosi}, {Rupchandani}, {Serrau}, {Shokry}, {Simon}, {Smith},
  {Sokoliuk}, {Soliman}, {Song}, {Takey}, {Tillayev}, {Tinjaca Ramirez}, {Tosta
  e Melo}, {Turpin}, {de Ugarte Postigo}, {Vanaverbeke}, {Vasylenko}, {Vernet},
  {Vidadi}, {Wang}, {Wang}, {Wang}, {Wang}, {Xiong}, {Xu}, {Xue}, {Zeng},
  {Zhang}, {Zhao}, \& {Zhao}}]{2023arXiv230206225K}
{Kann}, D.~A., {Agayeva}, S., {Aivazyan}, V., {et~al.} 2023, arXiv e-prints,
  arXiv:2302.06225, \dodoi{10.48550/arXiv.2302.06225}

\bibitem[{{Kippen} {et~al.}(1998){Kippen}, {Briggs}, {Kommers}, {Kouveliotou},
  {Hurley}, {Robinson}, {van Paradijs}, {Hartmann}, {Galama}, \&
  {Vreeswijk}}]{1998ApJ...506L..27K}
{Kippen}, R.~M., {Briggs}, M.~S., {Kommers}, J.~M., {et~al.} 1998, \apjl, 506,
  L27, \dodoi{10.1086/311634}

\bibitem[{{Klose} {et~al.}(2019){Klose}, {Schmidl}, {Kann}, {Nicuesa
  Guelbenzu}, {Schulze}, {Greiner}, {Olivares E.}, {Kr{\"u}hler}, {Schady},
  {Afonso}, {Filgas}, {Fynbo}, {Rau}, {Rossi}, {Takats}, {Tanga}, {Updike}, \&
  {Varela}}]{2019A&A...622A.138K}
{Klose}, S., {Schmidl}, S., {Kann}, D.~A., {et~al.} 2019, \aap, 622, A138,
  \dodoi{10.1051/0004-6361/201832728}

\bibitem[{Kn{\"o}dlseder(2020)}]{knodlseder2020cherenkov}
Kn{\"o}dlseder, J. 2020, arXiv preprint arXiv:2004.09213

\bibitem[{{Kraichnan}(1965)}]{1965PhFl....8.1385K}
{Kraichnan}, R.~H. 1965, Physics of Fluids, 8, 1385, \dodoi{10.1063/1.1761412}

\bibitem[{{Kuin} {et~al.}(2022){Kuin}, {Dichiara}, \& {Swift/UVOT
  Team}}]{2022GCN.32656....1K}
{Kuin}, N.~P.~M., {Dichiara}, S., \& {Swift/UVOT Team}. 2022, GRB Coordinates
  Network, 32656, 1

\bibitem[{{Kumar} \& {Zhang}(2015)}]{2015PhR...561....1K}
{Kumar}, P., \& {Zhang}, B. 2015, \physrep, 561, 1,
  \dodoi{10.1016/j.physrep.2014.09.008}

\bibitem[{{Laskar} {et~al.}(2015){Laskar}, {Berger}, {Margutti}, {Perley},
  {Zauderer}, {Sari}, \& {Fong}}]{2015ApJ...814....1L}
{Laskar}, T., {Berger}, E., {Margutti}, R., {et~al.} 2015, \apj, 814, 1,
  \dodoi{10.1088/0004-637X/814/1/1}

\bibitem[{{Laskar} {et~al.}(2023{\natexlab{a}}){Laskar}, {Alexander},
  {Margutti}, {Eftekhari}, {Chornock}, {Berger}, {Cendes}, {Duerr}, {Perley},
  {Edvige Ravasio}, {Yamazaki}, {Ayache}, {Barclay}, {Barniol Duran},
  {Bhandari}, {Brethauer}, {Christy}, {Coppejans}, {Duffell}, {Fong}, {Gomboc},
  {Guidorzi}, {Kennea}, {Kobayashi}, {Levan}, {Lobanov}, {Metzger}, {Ros},
  {Schroeder}, \& {Williams}}]{2023arXiv230204388L}
{Laskar}, T., {Alexander}, K.~D., {Margutti}, R., {et~al.} 2023{\natexlab{a}},
  arXiv e-prints, arXiv:2302.04388, \dodoi{10.48550/arXiv.2302.04388}

\bibitem[{{Laskar} {et~al.}(2023{\natexlab{b}}){Laskar}, {Alexander},
  {Margutti}, {Eftekhari}, {Chornock}, {Berger}, {Cendes}, {Duerr}, {Perley},
  {Ravasio}, {Yamazaki}, {Ayache}, {Barclay}, {Barniol Duran}, {Bhandari},
  {Brethauer}, {Christy}, {Coppejans}, {Duffell}, {Fong}, {Gomboc}, {Guidorzi},
  {Kennea}, {Kobayashi}, {Levan}, {Lobanov}, {Metzger}, {Ros}, {Schroeder}, \&
  {Williams}}]{2023ApJ...946L..23L}
---. 2023{\natexlab{b}}, \apjl, 946, L23, \dodoi{10.3847/2041-8213/acbfad}

\bibitem[{{Lemoine} \& {Revenu}(2006)}]{2006MNRAS.366..635L}
{Lemoine}, M., \& {Revenu}, B. 2006, \mnras, 366, 635,
  \dodoi{10.1111/j.1365-2966.2005.09912.x}

\bibitem[{{Lesage} {et~al.}(2022){Lesage}, {Veres}, {Roberts}, {Burns},
  {Bissaldi}, \& {Fermi GBM Team}}]{2022GCN.32642....1L}
{Lesage}, S., {Veres}, P., {Roberts}, O.~J., {et~al.} 2022, GRB Coordinates
  Network, 32642, 1

\bibitem[{{Lesage} {et~al.}(2023){Lesage}, {Veres}, {Briggs}, {Goldstein},
  {Kocevski}, {Burns}, {Wilson-Hodge}, {Bhat}, {Huppenkothen}, {Fryer},
  {Hamburg}, {Racusin}, {Bissaldi}, {Cleveland}, {Dalessi}, {Fletcher},
  {Giles}, {Hristov}, {Hui}, {Mailyan}, {Poolakkil}, {Roberts}, {von Kienlin},
  {Wood}, {Ajello}, {Arimoto}, {Baldini}, {Ballet}, {Baring}, {Bastieri},
  {Becerra Gonzalez}, {Bellazzini}, {Bissaldi}, {Blandford}, {Bonino}, {Bruel},
  {Buson}, {Cameron}, {Caputo}, {Caraveo}, {Cavazzuti}, {Chiaro}, {Cibrario},
  {Ciprini}, {Cristarella Orestano}, {Crnogorcevic}, {Cuoco}, {Cutini},
  {DAmmando}, {De Gaetano}, {Di Lalla}, {Di Venere}, {Dominguez}, {Fegan},
  {Ferrara}, {Fleischhack}, {Fukazawa}, {Funk}, {Fusco}, {Galanti}, {Gammaldi},
  {Gargano}, {Gasbarra}, {Gasparrini}, {Germani}, {Giacchino}, {Giglietto},
  {Gill}, {Giroletti}, {Granot}, {Green}, {Grenier}, {Guiriec}, {Gustafsson},
  {Hays}, {Hewitt}, {Horan}, {Hou}, {Kuss}, {Latronico}, {Laviron},
  {Lemoine-Goumard}, {Li}, {Liodakis}, {Longo}, {Loparco}, {Lorusso},
  {Lovellette}, {Lubrano}, {Maldera}, {Manfreda}, {Marti-Devesa}, {Mazziotta},
  {McEnery}, {Mereu}, {Meyer}, {Michelson}, {Mizuno}, {Monzani}, {Morselli},
  {Moskalenko}, {Negro}, {Nuss}, {Omodei}, {Orlando}, {Ormes}, {Paneque},
  {Panzarini}, {Persic}, {Pesce-Rollins}, {Pillera}, {Piron}, {Poon}, {Porter},
  {Principe}, {Raino}, {Rando}, {Rani}, {Razzano}, {Razzaque}, {Reimer},
  {Reimer}, {Ryde}, {Sanchez-Conde}, {Saz Parkinson}, {Scotton}, {Serini},
  {Sgro}, {Sharma}, {Siskind}, {Spandre}, {Spinelli}, {Tajima}, {Torres},
  {Valverde}, {Venters}, {Wadiasingh}, {Wood}, \&
  {Zaharijas}}]{2023arXiv230314172L}
{Lesage}, S., {Veres}, P., {Briggs}, M.~S., {et~al.} 2023, arXiv e-prints,
  arXiv:2303.14172, \dodoi{10.48550/arXiv.2303.14172}

\bibitem[{LHAASO-Collaboration {et~al.}(2023)LHAASO-Collaboration, Cao,
  Aharonian, An, null, Bai, Bai, Bao, Bastieri, Bi, Bi, Cai, Cao, Cao, Cao,
  Chang, Chang, Chen, Chen, Chen, Chen, Chen, Chen, Chen, Chen, Chen, Chen,
  Chen, Cheng, Cheng, Cheng, Cui, Cui, Cui, Dai, Dai, Dai, null, della Volpe,
  Dong, Duan, Fan, Fan, Fang, Fang, Feng, Feng, Feng, Feng, Feng, Gao, Gao,
  Gao, Gao, Gao, Gao, Ge, Geng, Gong, Gou, Gu, Guo, Guo, Guo, Guo, Han, He, He,
  He, He, He, Heller, Hor, Hou, Hou, Hou, Hu, Hu, Hu, Huang, Huang, Huang,
  Huang, Huang, Huang, Huang, Ji, Jia, Jia, Jiang, Jiang, Jiang, Jin, Kang, Ke,
  Kuleshov, Kurinov, Li, Li, Li, Li, Li, Li, Li, Li, Li, Li, Li, Li, Li, Li,
  Li, Li, Li, Li, Li, Liang, Liang, Lin, Liu, Liu, Liu, Liu, Liu, Liu, Liu,
  Liu, Liu, Liu, Liu, Liu, Liu, Liu, Liu, Liu, Long, Lu, Luo, Lv, Ma, Ma, Ma,
  Mao, Min, Mitthumsiri, Nan, Ou, Pang, Pattarakijwanich, Pei, Qi, Qi, Qiao,
  Qin, Ruffolo, Sáiz, Shao, Shao, Shchegolev, Sheng, Song, Stenkin, Stepanov,
  Su, Sun, Sun, Sun, Tam, Tang, Tian, Wang, Wang, Wang, Wang, Wang, Wang, Wang,
  Wang, Wang, Wang, Wang, Wang, Wang, Wang, Wang, Wang, Wang, Wang, Wang, Wang,
  Wang, Wang, Wei, Wei, Wei, Wen, Wu, Wu, Wu, Wu, Wu, Xi, Xia, Xia, Xiang,
  Xiao, Xiao, Xin, Xin, Xing, Xiong, Xu, Xu, Xu, Xue, Yan, Yan, Yan, Yang,
  Yang, Yang, Yang, Yang, Yang, Yang, Yang, Yang, Yao, Yao, Ye, Yin, Yin, You,
  You, Yu, Yuan, Yue, Zeng, Zeng, Zeng, Zeng, Zha, Zhang, Zhang, Zhang, Zhang,
  Zhang, Zhang, Zhang, Zhang, Zhang, Zhang, Zhang, Zhang, Zhang, Zhang, Zhang,
  Zhang, Zhang, Zhang, Zhang, Zhao, Zhao, Zhao, Zhao, Zhao, Zheng, Zheng, Zhou,
  Zhou, Zhou, Zhou, Zhou, Zhou, Zhu, Zhu, Zhu, Zhu, \&
  Zuo}]{doi:10.1126/science.adg9328}
LHAASO-Collaboration, Cao, Z., Aharonian, F., {et~al.} 2023, Science, 0,
  eadg9328, \dodoi{10.1126/science.adg9328}

\bibitem[{Lorenz(2005)}]{lorenz_magic_2005}
Lorenz, Eckartand~Martinez, M. 2005, Astronomy \& Geophysics, 46, 6.21,
  \dodoi{10.1111/j.1468-4004.2005.46621.x}

\bibitem[{{MAGIC Collaboration} {et~al.}(2019){MAGIC Collaboration}, {Acciari},
  {Ansoldi}, {Antonelli}, {Arbet Engels}, {Baack}, {Babi{\'c}}, {Banerjee},
  {Barres de Almeida}, {Barrio}, {Becerra Gonz{\'a}lez}, {Bednarek},
  {Bellizzi}, {Bernardini}, {Berti}, {Besenrieder}, {Bhattacharyya},
  {Bigongiari}, {Biland}, {Blanch}, {Bonnoli}, {Bo{\v{s}}njak}, {Busetto},
  {Carosi}, {Carosi}, {Ceribella}, {Chai}, {Chilingaryan}, {Cikota}, {Colak},
  {Colin}, {Colombo}, {Contreras}, {Cortina}, {Covino}, {D'Amico}, {D'Elia},
  {da Vela}, {Dazzi}, {de Angelis}, {de Lotto}, {Delfino}, {Delgado},
  {Depaoli}, {di Pierro}, {di Venere}, {Do Souto Espi{\~n}eira}, {Dominis
  Prester}, {Donini}, {Dorner}, {Doro}, {Elsaesser}, {Fallah Ramazani},
  {Fattorini}, {Fern{\'a}ndez-Barral}, {Ferrara}, {Fidalgo}, {Foffano},
  {Fonseca}, {Font}, {Fruck}, {Fukami}, {Gallozzi}, {Garc{\'\i}a L{\'o}pez},
  {Garczarczyk}, {Gasparyan}, {Gaug}, {Giglietto}, {Giordano}, {Godinovi{\'c}},
  {Green}, {Guberman}, {Hadasch}, {Hahn}, {Herrera}, {Hoang}, {Hrupec},
  {H{\"u}tten}, {Inada}, {Inoue}, {Ishio}, {Iwamura}, {Jouvin}, {Kerszberg},
  {Kubo}, {Kushida}, {Lamastra}, {Lelas}, {Leone}, {Lindfors}, {Lombardi},
  {Longo}, {L{\'o}pez}, {L{\'o}pez-Coto}, {L{\'o}pez-Oramas}, {Loporchio},
  {Machado de Oliveira Fraga}, {Maggio}, {Majumdar}, {Makariev}, {Mallamaci},
  {Maneva}, {Manganaro}, {Mannheim}, {Maraschi}, {Mariotti}, {Mart{\'\i}nez},
  {Masuda}, {Mazin}, {Mi{\'c}anovi{\'c}}, {Miceli}, {Minev}, {Miranda},
  {Mirzoyan}, {Molina}, {Moralejo}, {Morcuende}, {Moreno}, {Moretti},
  {Munar-Adrover}, {Neustroev}, {Nigro}, {Nilsson}, {Ninci}, {Nishijima},
  {Noda}, {Nogu{\'e}s}, {N{\"o}the}, {Nozaki}, {Paiano}, {Palacio},
  {Palatiello}, {Paneque}, {Paoletti}, {Paredes}, {Pe{\~n}il}, {Peresano},
  {Persic}, {Prada Moroni}, {Prandini}, {Puljak}, {Rhode}, {Rib{\'o}}, {Rico},
  {Righi}, {Rugliancich}, {Saha}, {Sahakyan}, {Saito}, {Sakurai}, {Satalecka},
  {Schmidt}, {Schweizer}, {Sitarek}, {{\v{S}}nidari{\'c}}, {Sobczynska},
  {Somero}, {Stamerra}, {Strom}, {Strzys}, {Suda}, {Suri{\'c}}, {Takahashi},
  {Tavecchio}, {Temnikov}, {Terzi{\'c}}, {Teshima}, {Torres-Alb{\`a}}, {Tosti},
  {Tsujimoto}, {Vagelli}, {van Scherpenberg}, {Vanzo}, {Vazquez Acosta},
  {Vigorito}, {Vitale}, {Vovk}, {Will}, {Zari{\'c}}, \&
  {Nava}}]{2019Natur.575..455M}
{MAGIC Collaboration}, {Acciari}, V.~A., {Ansoldi}, S., {et~al.} 2019, \nat,
  575, 455, \dodoi{10.1038/s41586-019-1750-x}

\bibitem[{{Medvedev}(2006)}]{2006ApJ...651L...9M}
{Medvedev}, M.~V. 2006, \apjl, 651, L9, \dodoi{10.1086/509075}

\bibitem[{{Medvedev} \& {Loeb}(1999)}]{1999ApJ...526..697M}
{Medvedev}, M.~V., \& {Loeb}, A. 1999, \apj, 526, 697, \dodoi{10.1086/308038}

\bibitem[{{M{\'e}sz{\'a}ros}(2006)}]{2006RPPh...69.2259M}
{M{\'e}sz{\'a}ros}, P. 2006, Reports on Progress in Physics, 69, 2259,
  \dodoi{10.1088/0034-4885/69/8/R01}

\bibitem[{{M{\'e}sz{\'a}ros} {et~al.}(1998){M{\'e}sz{\'a}ros}, {Rees}, \&
  {Wijers}}]{1998ApJ...499..301M}
{M{\'e}sz{\'a}ros}, P., {Rees}, M.~J., \& {Wijers}, R.~A.~M.~J. 1998, \apj,
  499, 301, \dodoi{10.1086/305635}

\bibitem[{{Miceli} \& {Nava}(2022)}]{2022Galax..10...66M}
{Miceli}, D., \& {Nava}, L. 2022, Galaxies, 10, 66,
  \dodoi{10.3390/galaxies10030066}

\bibitem[{{Nava}(2018)}]{2018IJMPD..2742003N}
{Nava}, L. 2018, International Journal of Modern Physics D, 27, 1842003,
  \dodoi{10.1142/S0218271818420038}

\bibitem[{{Niemiec} {et~al.}(2006){Niemiec}, {Ostrowski}, \&
  {Pohl}}]{2006ApJ...650.1020N}
{Niemiec}, J., {Ostrowski}, M., \& {Pohl}, M. 2006, \apj, 650, 1020,
  \dodoi{10.1086/506901}

\bibitem[{{O'Connor} {et~al.}(2023){O'Connor}, {Troja}, {Ryan}, {Beniamini},
  {van Eerten}, {Granot}, {Dichiara}, {Ricci}, {Lipunov}, {Gillanders}, {Gill},
  {Moss}, {Anand}, {Andreoni}, {Becerra}, {Buckley}, {Butler}, {Cenko},
  {Chasovnikov}, {Durbak}, {Francile}, {Hammerstein}, {van der Horst},
  {Kasliwal}, {Kouveliotou}, {Kutyrev}, {Lee}, {Srinivasaragavan}, {Topolev},
  {Watson}, {Yang}, \& {Zhirkov}}]{2023arXiv230207906O}
{O'Connor}, B., {Troja}, E., {Ryan}, G., {et~al.} 2023, arXiv e-prints,
  arXiv:2302.07906, \dodoi{10.48550/arXiv.2302.07906}

\bibitem[{Omodei.~N.(2022{\natexlab{a}})}]{web:nasa:LAT1}
Omodei.~N., bruel.~P., . b. J. e.~a. 2022{\natexlab{a}}, GRB 221009A: Fermi LAT
  data rate effects due to extremely high flux.
\newblock \url{https://gcn.gsfc.nasa.gov/gcn3/32760.gcn3}

\bibitem[{Omodei.~N.(2022{\natexlab{b}})}]{web:nasa:LAT2}
---. 2022{\natexlab{b}}, GRB 221009A: Extended Bad Time Intervals for Fermi LAT
  data.
\newblock \url{https://gcn.gsfc.nasa.gov/gcn3/32916.gcn3}

\bibitem[{{Paczynski}(1990)}]{1990ApJ...348..485P}
{Paczynski}, B. 1990, \apj, 348, 485, \dodoi{10.1086/168257}

\bibitem[{{Paczynski} \& {Rhoads}(1993)}]{1993ApJ...418L...5P}
{Paczynski}, B., \& {Rhoads}, J.~E. 1993, \apjl, 418, L5,
  \dodoi{10.1086/187102}

\bibitem[{{Panaitescu} \& {Kumar}(2000)}]{2000ApJ...543...66P}
{Panaitescu}, A., \& {Kumar}, P. 2000, \apj, 543, 66, \dodoi{10.1086/317090}

\bibitem[{Pillera {et~al.}(2022)Pillera, Bissaldi, \&
  Omodei}]{web:nasa:FermilAT}
Pillera, R., Bissaldi, E., \& Omodei, N., e.~a. 2022, GRB 221009A: Fermi-LAT
  refined analysis.
\newblock \url{https://gcn.gsfc.nasa.gov/gcn3/32658.gcn3}

\bibitem[{{Piran}(1999)}]{1999PhR...314..575P}
{Piran}, T. 1999, \physrep, 314, 575, \dodoi{10.1016/S0370-1573(98)00127-6}

\bibitem[{Piran {et~al.}(1993)Piran, Shemi, \& Narayan}]{Piran:1993b}
Piran, T., Shemi, A., \& Narayan, R. 1993, Monthly Notices of the Royal
  Astronomical Society, 263, 861, \dodoi{10.1093/mnras/263.4.861}

\bibitem[{{Razzaque} {et~al.}(2010){Razzaque}, {Dermer}, \&
  {Finke}}]{2010OAJ.....3..150R}
{Razzaque}, S., {Dermer}, C.~D., \& {Finke}, J.~D. 2010, The Open Astronomy
  Journal, 3, 150, \dodoi{10.2174/1874381101003010150}

\bibitem[{{Ren} {et~al.}(2022){Ren}, {Wang}, \& {Zhang}}]{2022arXiv221010673R}
{Ren}, J., {Wang}, Y., \& {Zhang}, L.-L. 2022, arXiv e-prints,
  arXiv:2210.10673, \dodoi{10.48550/arXiv.2210.10673}

\bibitem[{{Ressler} \& {Laskar}(2017)}]{2017ApJ...845..150R}
{Ressler}, S.~M., \& {Laskar}, T. 2017, \apj, 845, 150,
  \dodoi{10.3847/1538-4357/aa8268}

\bibitem[{{Reynolds} {et~al.}(2021){Reynolds}, {Williams}, {Borkowski}, \&
  {Long}}]{2021ApJ...917...55R}
{Reynolds}, S.~P., {Williams}, B.~J., {Borkowski}, K.~J., \& {Long}, K.~S.
  2021, \apj, 917, 55, \dodoi{10.3847/1538-4357/ac0ced}

\bibitem[{{Rybicki} \& {Lightman}(1986)}]{1986rpa..book.....R}
{Rybicki}, G.~B., \& {Lightman}, A.~P. 1986, {Radiative Processes in
  Astrophysics} (Wiley-VCH)

\bibitem[{{Salafia} {et~al.}(2022){Salafia}, {Ravasio}, {Yang}, {An},
  {Orienti}, {Ghirlanda}, {Nava}, {Giroletti}, {Mohan}, {Spinelli}, {Zhang},
  {Marcote}, {Cim{\`o}}, {Wu}, \& {Li}}]{2022ApJ...931L..19S}
{Salafia}, O.~S., {Ravasio}, M.~E., {Yang}, J., {et~al.} 2022, \apjl, 931, L19,
  \dodoi{10.3847/2041-8213/ac6c28}

\bibitem[{{Sari} \& {Esin}(2001)}]{2001ApJ...548..787S}
{Sari}, R., \& {Esin}, A.~A. 2001, \apj, 548, 787, \dodoi{10.1086/319003}

\bibitem[{{Sari} \& {Piran}(1995)}]{Sari:1995ApJ}
{Sari}, R., \& {Piran}, T. 1995, \apjl, 455, L143, \dodoi{10.1086/309835}

\bibitem[{Sari {et~al.}(1998)Sari, Piran, \& Narayan}]{Sari_1998}
Sari, R., Piran, T., \& Narayan, R. 1998, The Astrophysical Journal, 497, L17,
  \dodoi{10.1086/311269}

\bibitem[{{Sato} {et~al.}(2023){Sato}, {Murase}, {Ohira}, \&
  {Yamazaki}}]{2023MNRAS.522L..56S}
{Sato}, Y., {Murase}, K., {Ohira}, Y., \& {Yamazaki}, R. 2023, \mnras, 522,
  L56, \dodoi{10.1093/mnrasl/slad038}

\bibitem[{{Schlegel} {et~al.}(1998){Schlegel}, {Finkbeiner}, \&
  {Davis}}]{1998ApJ...500..525S}
{Schlegel}, D.~J., {Finkbeiner}, D.~P., \& {Davis}, M. 1998, \apj, 500, 525,
  \dodoi{10.1086/305772}

\bibitem[{{Shrestha} {et~al.}(2023){Shrestha}, {Sand}, {Alexander}, {Bostroem},
  {Hosseinzadeh}, {Pearson}, {Aghakhanloo}, {Vink{\'o}}, {Andrews}, {Jencson},
  {Lundquist}, {Wyatt}, {Howell}, {McCully}, {Gonzalez}, {Pellegrino},
  {Terreran}, {Hiramatsu}, {Newsome}, {Farah}, {Jha}, {Smith}, {Wheeler},
  {Mart{\'\i}nez-V{\'a}zquez}, {Carballo-Bello}, {Drlica-Wagner}, {James},
  {Mutlu-Pakdil}, {Stringfellow}, {Sakowska}, {No{\"e}l}, {Bom}, \&
  {Kuehn}}]{2023ApJ...946L..25S}
{Shrestha}, M., {Sand}, D.~J., {Alexander}, K.~D., {et~al.} 2023, \apjl, 946,
  L25, \dodoi{10.3847/2041-8213/acbd50}

\bibitem[{{Sironi} \& {Spitkovsky}(2011)}]{2011ApJ...726...75S}
{Sironi}, L., \& {Spitkovsky}, A. 2011, \apj, 726, 75,
  \dodoi{10.1088/0004-637X/726/2/75}

\bibitem[{{Tan} {et~al.}(2022){Tan}, {Li}, {Ge}, {Li}, {Xiong}, \&
  {Zhang}}]{2022ATel15660....1T}
{Tan}, W.~J., {Li}, C.~K., {Ge}, M.~Y., {et~al.} 2022, The Astronomer's
  Telegram, 15660, 1

\bibitem[{{Totani}(1998)}]{1998ApJ...totania}
{Totani}, T. 1998, \apjl, 502, L13, \dodoi{10.1086/311489}

\bibitem[{Veres {et~al.}(2022)Veres, Burns, \& Bissaldi}]{web:nasa:FermiGBM}
Veres, P., Burns, E., \& Bissaldi, E., e.~a. 2022, GRB 221009A: Fermi GBM
  detection of an extraordinarily bright GRB.
\newblock \url{https://gcn.gsfc.nasa.gov/gcn3/32636.gcn3}

\bibitem[{Vietri(1997)}]{PhysRevLett.78.4328}
Vietri, M. 1997, Phys. Rev. Lett., 78, 4328,
  \dodoi{10.1103/PhysRevLett.78.4328}

\bibitem[{{Wang} {et~al.}(2019){Wang}, {Liu}, {Zhang}, {Xi}, \&
  {Zhang}}]{2019ApJ...884..117W}
{Wang}, X.-Y., {Liu}, R.-Y., {Zhang}, H.-M., {Xi}, S.-Q., \& {Zhang}, B. 2019,
  \apj, 884, 117, \dodoi{10.3847/1538-4357/ab426c}

\bibitem[{{Warren} {et~al.}(2018){Warren}, {Barkov}, {Ito}, {Nagataki}, \&
  {Laskar}}]{2018MNRAS.480.4060W}
{Warren}, D.~C., {Barkov}, M.~V., {Ito}, H., {Nagataki}, S., \& {Laskar}, T.
  2018, \mnras, 480, 4060, \dodoi{10.1093/mnras/sty2138}

\bibitem[{{Waxman} \& {Bahcall}(2000)}]{2000ApJ...541..707W}
{Waxman}, E., \& {Bahcall}, J.~N. 2000, \apj, 541, 707, \dodoi{10.1086/309462}

\bibitem[{{Williams} {et~al.}(2023){Williams}, {Kennea}, {Dichiara},
  {Kobayashi}, {Iwakiri}, {Beardmore}, {Evans}, {Heinz}, {Lien}, {Oates},
  {Negoro}, {Cenko}, {Buisson}, {Hartmann}, {Jaisawal}, {Kuin}, {Lesage},
  {Page}, {Parsotan}, {Pasham}, {Sbarufatti}, {Siegel}, {Sugita}, {Younes},
  {Ambrosi}, {Arzoumanian}, {Bernardini}, {Campana}, {Capalbi}, {Caputo},
  {D'A{\`\i}}, {D'Avanzo}, {D'Elia}, {De Pasquale}, {Eyles-Ferris}, {Ferrara},
  {Gendreau}, {Gropp}, {Kawai}, {Klingler}, {Laha}, {Melandri}, {Mihara},
  {Moss}, {O'Brien}, {Osborne}, {Palmer}, {Perri}, {Serino}, {Sonbas},
  {Stamatikos}, {Starling}, {Tagliaferri}, {Tohuvavohu}, {Zane}, \&
  {Ziaeepour}}]{2023ApJ...946L..24W}
{Williams}, M.~A., {Kennea}, J.~A., {Dichiara}, S., {et~al.} 2023, \apjl, 946,
  L24, \dodoi{10.3847/2041-8213/acbcd1}

\bibitem[{{Woosley} {et~al.}(2002){Woosley}, {Heger}, \&
  {Weaver}}]{2002RvMP...74.1015W}
{Woosley}, S.~E., {Heger}, A., \& {Weaver}, T.~A. 2002, Reviews of Modern
  Physics, 74, 1015, \dodoi{10.1103/RevModPhys.74.1015}

\bibitem[{{Wright}(2006)}]{2006PASP..118.1711W}
{Wright}, E.~L. 2006, \pasp, 118, 1711, \dodoi{10.1086/510102}

\bibitem[{{Yamasaki} \& {Piran}(2022)}]{2022MNRAS.512.2142Y}
{Yamasaki}, S., \& {Piran}, T. 2022, \mnras, 512, 2142,
  \dodoi{10.1093/mnras/stac483}

\bibitem[{Zhang(2018)}]{zhang_2018}
Zhang, B. 2018, The Physics of Gamma-Ray Bursts (Cambridge University Press),
  \dodoi{10.1017/9781139226530}

\bibitem[{{Zhang} \& {M{\'e}sz{\'a}ros}(2001)}]{2001ApJ...559..110Z}
{Zhang}, B., \& {M{\'e}sz{\'a}ros}, P. 2001, \apj, 559, 110,
  \dodoi{10.1086/322400}

\bibitem[{{Zhang} {et~al.}(2007){Zhang}, {Liang}, {Page}, {Grupe}, {Zhang},
  {Barthelmy}, {Burrows}, {Campana}, {Chincarini}, {Gehrels}, {Kobayashi},
  {M{\'e}sz{\'a}ros}, {Moretti}, {Nousek}, {O'Brien}, {Osborne}, {Roming},
  {Sakamoto}, {Schady}, \& {Willingale}}]{2007ApJ...655..989Z}
{Zhang}, B., {Liang}, E., {Page}, K.~L., {et~al.} 2007, \apj, 655, 989,
  \dodoi{10.1086/510110}

\bibitem[{{Zhang} {et~al.}(2022){Zhang}, {Murase}, {Ioka}, {Song}, {Yuan}, \&
  {M{\'e}sz{\'a}ros}}]{2022arXiv221105754Z}
{Zhang}, B.~T., {Murase}, K., {Ioka}, K., {et~al.} 2022, arXiv e-prints,
  arXiv:2211.05754, \dodoi{10.48550/arXiv.2211.05754}

\bibitem[{{Zhang} {et~al.}(2018){Zhang}, {Murase}, {Kimura}, {Horiuchi}, \&
  {M{\'e}sz{\'a}ros}}]{2018PhRvD..97h3010Z}
{Zhang}, B.~T., {Murase}, K., {Kimura}, S.~S., {Horiuchi}, S., \&
  {M{\'e}sz{\'a}ros}, P. 2018, \prd, 97, 083010,
  \dodoi{10.1103/PhysRevD.97.083010}

\end{thebibliography}
\bibliographystyle{aasjournal}



\end{document}